\documentclass[reqno,11pt]{amsart}
\usepackage{graphicx}
\usepackage{amscd}
\usepackage{slashed}
\usepackage{amssymb}
\usepackage[mathscr]{eucal}
\numberwithin{equation}{section}
\allowdisplaybreaks[1]

\newcommand{\SetFigFont}[3]{}

\title[Gupta-Bleuler Quantization in Globally Hyperbolic Space-Times]
{Gupta-Bleuler Quantization of the Maxwell Field in Globally Hyperbolic Space-Times}

\author[F.\ Finster]{Felix Finster}
\address{Fakult\"at f\"ur Mathematik \\ Universit\"at Regensburg \\ D-93040 Regensburg \\ Germany}
\email{finster@ur.de}
\thanks{F.F.\ was supported by the Scheme 4 Grant~\#41122 of the London Mathematical Society}

\author[A.\ Strohmaier]{Alexander Strohmaier \\ \\ July 2020}
\address{Department of Mathematical Sciences \\ Loughborough University \\
Loughborough, LE11 3TU, United Kingdom}
\email{a.strohmaier@lboro.ac.uk}

\newtheorem{Def}{Definition}[section]
\newtheorem{Thm}[Def]{Theorem}
\newtheorem{Prp}[Def]{Proposition}
\newtheorem{Lemma}[Def]{Lemma}
\newtheorem{Remark}[Def]{Remark}
\newtheorem{Corollary}[Def]{Corollary}

\newtheorem{Assumption}[Def]{Assumption}
\newcommand{\Thanks}{\vspace*{.5em} \noindent \thanks}
\newcommand{\beq}{\begin{equation}}
\newcommand{\eeq}{\end{equation}}
\newcommand{\Proof}{\begin{proof}}
\newcommand{\QED}{\end{proof} \noindent}
\newcommand{\QEDrem}{\ \hfill $\Diamond$}

\newcommand{\la}{\langle}
\newcommand{\ra}{\rangle}

\newcommand{\C}{\mathbb{C}}
\newcommand{\R}{\mathbb{R}}
\newcommand{\1}{\mbox{\rm 1 \hspace{-1.05 em} 1}}

\newcommand{\N}{\mathbb{N}}

\renewcommand{\H}{\mathscr{H}}

\newcommand{\f}{{\mathfrak{f}}}
\newcommand{\g}{{\mathfrak{g}}}

\newcommand{\bep}{\begin{pmatrix}}
\newcommand{\enp}{\end{pmatrix}}

\newcommand{\D}{{\mathscr{D}}}
\newcommand{\K}{{\mathcal{K}}}

\renewcommand{\L}{{\mathcal{L}}}

\newcommand{\I}{{\mathfrak{I}}}

\DeclareMathOperator{\im}{Im}

\DeclareMathOperator{\supp}{supp}

\DeclareMathOperator{\End}{\mbox{\rm{End}}}

\newcommand{\rmi}{i} 
\newcommand{\PDO}{\psi {\rm{DO}}}
\newcommand{\WF}{{\rm{WF}}}
\newcommand{\WFpol}{{\rm{WF}}_{\!\text{\tiny{\rm{pol}}}}}

\begin{document}
\maketitle

\begin{abstract}
We give a complete framework for the Gupta-Bleuler quantization of the free electromagnetic
field on globally hyperbolic space-times.
We describe one-particle structures that give rise to states satisfying the microlocal spectrum
condition. The field algebras in the so-called Gupta-Bleuler representations 
satisfy the time-slice axiom, and the corresponding vacuum states satisfy the microlocal spectrum condition.
We also give an explicit construction of ground states
on ultrastatic space-times. Unlike previous constructions, our method does not require 
a spectral gap or the absence of zero modes. The only requirement, the absence of zero-resonance states,
is shown to be stable under compact perturbations of topology and metric. Usual deformation arguments
based on the time-slice axiom then lead to a construction of Gupta-Bleuler representations on a large
class of globally hyperbolic space-times.
As usual, the field algebra is represented on an indefinite inner product space, in which the
physical states form a positive semi-definite subspace. Gauge transformations are incorporated
in such a way that the field can be coupled perturbatively to a Dirac field. Our approach does not require any
topological restrictions on the underlying space-time.
\end{abstract}

\tableofcontents

\section{Introduction}
The classical approach to perturbative quantum electrodynamics
begins with the canonical quantization scheme for the Dirac field and the electromagnetic field. 
While the free electromagnetic field may well be described by the
quantized field algebra of the electromagnetic field strength, minimal coupling to a Dirac field
requires the quantization of the vector potential. In a coupled theory, the gauge group will act
on both the Dirac field and the vector potential. Therefore, the gauge group cannot be factored out
before the coupling is introduced.
It was realized by Gupta and Bleuler~\cite{gupta, bleuleralt} that in Minkowski space-time,
the field algebra of the vector potential is most conveniently represented in a Poincar\'e covariant
manner on an indefinite inner product space in which the physical states form a positive semi-definite
subspace. Perturbative quantum electrodynamics can then be carried out consistently on the level of
formal power series.

Conceptually, it is best to split the construction of
the field operators in canonical quantization into two steps. Step one is to construct the so called field algebra,
i.e.\ a $*$-algebra that satisfies the canonical commutation relations. Step two consists in finding
representations of this algebra that are physically reasonable.
In Minkowski space-time, there is usually
a preferred Poincar\'e invariant ground state and therefore a physically preferred representation.
In this situation, the construction is canonical and is often carried out in one step by employing a
procedure which in physics is called frequency splitting.

In quantum field theory on curved space-time, one considers quantized fields on a classical curved space-time.
In a space-time~$(M^n,g)$, the classical Maxwell equations can be formulated
with differential forms by
\[ dF =0,\qquad \delta F = J\:, \]
where~$F \in \Omega^2(M)$ is the field strength, and~$J$ is the electromagnetic current.
In the so-called potential method for finding cohomologically trivial solutions of these equations, one sets~$F=dA$
with a one-form~$A$, the so-called {\em{electromagnetic potential}}.
Then the equation~$dF=0$ is automatically satisfied, so that Maxwell's equations reduce to
\beq \label{maxa}
\delta d A = J \:.
\eeq
The potential~$A$ is not uniquely determined. Namely, transforming~$A$ according to
\beq \label{classgauge}
A(x) \mapsto A(x) - d \Lambda(x)
\eeq
with~$\Lambda \in \Omega^0(M)$ maps the solution space of~\eqref{maxa} to itself
and leaves~$F$ unchanged.
The transformations~\eqref{classgauge} are the {\em{classical gauge transformations}}
of electrodynamics.

The physical requirement of {\em{gauge invariance}} states that all observable quantities
should be invariant under gauge transformations.
In particular, the electromagnetic potential~$A$ is not gauge invariant.
The field strength~$F=dA$ is, making the electromagnetic field an observable quantity.
Another way of forming gauge invariant quantities is to integrate the electromagnetic potential
along a closed curve, or more generally a cycle,
\[ \int_\gamma A  \qquad \text{for a cycle $\gamma$}\:. \]
By Stokes' theorem, knowing $\int_\gamma A$ for all homologically trivial cycles
is equivalent to knowing the field strength $F=dA$. However, as the Aharanov-Bohm experiment shows
(see for example~\cite{peskin+tonomura}),
also homologically non-trivial cycles correspond to measurable quantities.
Thus not the field strength alone, but the integrals of the electromagnetic potential along
all cycles, should be regarded as the fundamental physical objects of electrodynamics.

This physical significance of the electromagnetic potential becomes clearer in
quantum mechanics, where~$A$ is needed to describe the coupling of the electromagnetic
field to the quantum mechanical particle. For example, in the Dirac equation the coupling
is described by the term~$\gamma^j A_j(x) \psi(x)$, which can be generated by the so-called
{\em{minimal coupling}} procedure where one replaces the partial derivatives in the Dirac equation
according to~$\partial_j \rightarrow \partial_j - i A_j(x)$.
Thus it is impossible to work with the field
strength alone; one must consider the potential~$A(x)$ as being the basic object
describing the electromagnetic field. In such a coupled situation, the gauge
transformations~\eqref{classgauge} extend to transformations on the whole system,
which typically describe local phase transformations of the wave functions
\[ \psi(x) \mapsto e^{-i \Lambda(x)}\: \psi(x)\:. \]
In geometric terms, minimal coupling is best understood as follows. The solutions of the Dirac equation
are sections of a Dirac bundle that is twisted by a line bundle. The classical electromagnetic field
vector potential should be regarded as a connection on this line bundle, and the Dirac equation
is formed using the connection on the bundle. Once a local trivialization of the line bundle and the Dirac
bundle is fixed, the connection determines a one-form, the vector potential. Gauge transformations then correspond to different choices of local trivializations.

As a consequence of the classical gauge freedom~\eqref{classgauge}, the
Cauchy problem for Maxwell's equations~\eqref{maxa} is ill-posed.
In order to circumvent this problem, one typically chooses a specific gauge. A common choice is
the {\em{Lorenz gauge}}
\beq \label{Lorenz}
\delta A = 0 \:.
\eeq
Then the Maxwell equations go over to the wave equation
\[ \square A = 0 \:. \]
When performing a gauge transformation~\eqref{classgauge}, the
gauge condition~\eqref{Lorenz} becomes
\beq \label{gaugec}
\delta A = \Box \Lambda\:,
\eeq
and the field equations transform to
\[ \square A = \square  d \Lambda \:. \]

The goal of this paper is to quantize the electromagnetic field in a curved space-time
in such a way that this field can readily be coupled to a quantized Dirac field.
We restrict attention to the first step where the electromagnetic field is quantized. 
In a second step, the coupling to quantum particles could be described
perturbatively.
With this in mind, we only consider the free dynamics for the vector potential without source term.
However, as the coupling to other particles and fields requires the electromagnetic potential,
we want to construct field operators~$\hat{A}$ for the electromagnetic potential.
We do not impose any cohomological restrictions on our space-time. When coupling to the Dirac field,
we assume that the quantization starts in a fixed topological sector. This means that we assume that
we have chosen a fixed line bundle and a fixed connection with respect to which we perturb. Thus
the electromagnetic potential to be quantized will consist of globally defined one-forms. Passing from one background connection to another
can be implemented by an algebra automorphism that may however not be unitarily implementable
in a given standard Fock representation (see e.g.~\cite{ashtekar+sen} and~\cite{streater}). Different background connections may therefore be thought of as choices of different classes
of representations or superselection sectors.

The paper is organized as follows. After a brief mathematical introduction
(Section~\ref{secprelim}), we define the field algebras, introduce gauge transformations
and prove the time slice axiom (Section~\ref{sec3}).
In Section~\ref{sec4}, we consider representations of the field algebras and explain the properties
which we demand from a physically reasonable representation.
More precisely, we define so-called {\em{Gupta-Bleuler representations}} as representations on an
indefinite inner product space which satisfy a microlocal spectrum condition.
Moreover, we demand that applying the observables to the vacuum should generate
the positive semi-definite subspace of physical states.
Furthermore, the gauge condition~$\delta \hat{A} = \Box \Lambda$ should be satisfied for the expectation
values of the physical states.
In Section~\ref{secultra}, we construct Gupta-Bleuler representations for ultrastatic manifolds.
Here our main point is to treat the zero resonance states (Section~\ref{secreson})
and the zero modes (Section~\ref{seckappa}).
Using a glueing construction, these representations are then extended to general
globally hyperbolic space-times (Section~\ref{sec8}).
All our constructions are gauge covariant, where we extend the classical gauge transformation
law~\eqref{classgauge} to the field operators~$\hat{A}$ by
\beq \label{fieldgauge2}
\hat{A}(x) \mapsto \hat{A}(x) - d \Lambda(x) \:,
\eeq
and~$\Lambda$ is again a real-valued function.
This corresponds to the usual procedure in canonical quantization schemes
(see for example~\cite[Section~8]{weinberg}) in which the gauge freedom 
described by so-called scalar photons is not quantized.
In particular, the gauge transformations~\eqref{fieldgauge2} leave the commutator relations
of the field operators unchanged. An alternative procedure described in the literature is to fix the gauge
with a {\em{gauge parameter}}. This leads to modifications of the commutator relations for non-observable
quantities. In Appendix~\ref{secgaugeparam} we show that 
working with different gauge parameters gives an equivalent description of the
physical system. Since the construction of one-particle states for massless spin one fields is interesting in its own
right and independent of the Gupta-Bleuler framework, we show in Appendix \ref{brst} that it can also be
used to construct states in the BRST framework as formulated in~\cite{duetsch+fredenhagen} and
in curved space-times in~\cite{hollands3}. In Appendix \ref{cohoint} it is shown that the presence of zero
modes is intimately related to cohomology and may be presented independent of foliations.

\subsection{Discussion and Relation to Previous Work}
In curved space-time, it was first shown by Dimock~\cite{dimock1}
that the algebra of the free scalar field on a globally hyperbolic space-time 
can be constructed in a functorial manner. Thus the first step, the construction of the
field algebra, can be carried out just as in Minkowski space-time.
Dimock later used this procedure to quantize the electromagnetic field strength~\cite{dimock2}.
The canonical quantization of the electromagnetic vector potential in a curved background in the
Gupta-Bleuler framework was first described by Furlani~\cite{furlani}, who assumes the space-time to
be ultrastatic with compact Cauchy surfaces. We note here, however, that in the presence of zero modes,
the construction given in~\cite{furlani} contains gaps (in particular, Theorem III.1 does not
hold if~$H^1(M)\not=\{0\}$, essentially because when projecting out the zero modes, the
locality of the commutation relations is lost).
Another series of papers~\cite{fewster+pfenning,dappiaggi, dappiaggi+siemssen} 
deals with the uncoupled electromagnetic field in curved space-times. More specifically,
in~\cite{fewster+pfenning} the field algebra for the field strength is constructed under certain cohomological conditions, and some representations are given.
In~\cite{dappiaggi} the field algebra for the field strength is constructed without
cohomological conditions. In the paper~\cite{dappiaggi+siemssen} the field algebra smeared out with
co-closed test functions is constructed; under the stated cohomological assumptions this algebra
is equivalent to the algebra for the field strength.
In the construction of physical representations, in previous works
the term ``Gupta-Bleuler quantization'' is ambiguous.
It is often referred to as a method to construct a state over the {\em{algebra of observables}} that satisfies
the microlocal spectrum condition. In the present paper, we take the view that such a quantization procedure 
should result in a physical representation of the {\em{whole field algebra}} that has all the properties needed
for coupling it to an electron field and for constructing quantum electrodynamics, at least on the level
of formal power series in a localized coupling constant.
The procedure we propose is as close as possible to the standard textbook Gupta-Bleuler construction 
in Minkowski space-time whilst at the same time keeping the language suitable for general globally hyperbolic
space-times and independent of preferred foliations. Our method of construction of Gupta-Bleuler states
in ultrastatic space-times does not require the presence of a spectral gap or the absence of zero modes. To
our knowledge, the construction in this generality is new even on the level of one-particle structures.
We therefore show in Appendix~\ref{brst} that it also applies to the BRST framework as described
in~\cite{henneaux, barnich, hollands3}.
Finally, the paper~\cite{dappiaggi+sanders} deals with the construction of the abstract
field algebras and observable algebras without topological restrictions and in the presence of classical
external sources, but physical representations of these algebras are not considered. Finally, a discussion of the
quantization of the full bundle of connections modulo gauge transformations can be found
in~\cite{benini1, benini2}.

\section{Mathematical Preliminaries} \label{secprelim}
Let~$(M, g)$ be a globally hyperbolic Lorentzian space-time of dimension~$n \geq 2$, i.e.\ $M$ is an oriented, 
time-oriented Lorentzian manifold that admits a smooth global Cauchy surface~$\Sigma$
(see~\cite{bernal+sanchez}). We assume that the metric has signature $(+1,-1,\ldots,-1)$. 
Let $\Omega^p(M) \subset C^\infty(M;\Lambda^p T^*M)$ be the space of smooth real-valued $p$-forms
and $\Omega^p_0(M) \subset \Omega^p(M)$ be the forms with compact support.
As usual denote by $d: \Omega^p(M) \to \Omega^{p+1}(M)$ the exterior derivative and by
$\delta: \Omega^{p+1}(M) \to \Omega^{p}(M)$ its formal adjoint with respect to the
inner product on the space of $p$-forms
$$ \langle f, g \rangle = \int_M f \wedge * g\:, $$
where $*$ is the Hodge star operator. Note that this inner product is indefinite if $0< p < n$.

The wave operator $\square_p : \Omega^p(M) \to \Omega^{p}(M)$ is defined
by~$\square_p=d \delta + \delta d$. It is formally self-adjoint with respect to the above inner product.
The wave equation $\square_p A =0$ for $p$-forms $A \in \Omega^p(M)$ is a normally hyperbolic  differential
equation. It is well-known that the Cauchy problem for this equation can be solved uniquely,
and moreover there exist unique advanced and retarded fundamental solutions~$G^p_{\pm} : \Omega^p_0(M)  \to  \Omega^p(M)$
such that
\begin{enumerate}
 \item $G^p_\pm$ is continuous with respect to the usual locally convex topologies on $\Omega^p_0(M)$  and  $\Omega^p(M)$, respectively.
 \item $\square_p G^p_{\pm} f = G^p_{\pm} \square_p f = f$ for all $f \in \Omega^p_0(M)$,
 \item $\supp G^p_\pm f \subset J^\pm(\supp f)$, where $J^\pm(\supp f)$ denotes the causal
 future respectively past of~$\supp f$.
\end{enumerate}
(we refer the reader to the monograph~\cite{baer+ginoux} for a detailed general proof in the context
of operators on vector bundles).
The map $G^p$ is then defined to be the difference of retarded and advanced
fundamental solutions~$G^p := G^p_+ - G^p_-$. Note that $G^p$ maps onto the space of smooth solutions
of the equation  $\square_p A =0$ with spatially compact support,
i.e.\ solutions whose support have compact intersection with $\Sigma$.
The function $G^p(f)$ can be viewed as a distribution and may be paired with a test function 
$g \in \Omega^p_0(M)$, using the inner product~$\la \cdot , \cdot \ra$.
We will denote $G^p(f)(g)
=  \la G^p(f), g \ra$ by $G^p(f,g)$. The bilinear form $G^p(\cdot,\cdot)$ defines a distribution
on $M \times M$ with values in $\Lambda^p T^*M \boxtimes \Lambda^p T^*M$. It is straightforward to verify that
$G^p(f,g) = - G^p(g,f)$.

Throughout the paper, we regard the space of $p$-forms as a subset of the set of distributional $p$-forms
by using the inner product. That is if $A \in \Omega^p(M)$ we may pair $A$ with a test function $f \in \Omega^p_0(M)$
$$
 A(f) := \int_M A \wedge * f.
$$
For example if $A$ is a one-form which in local coordinates is given by $A = \sum_{i=1}^n A_i(x) \,dx^i$,
then the corresponding integral in local coordinates is
$$
 A(f) = \int_M \bigg( \sum_{i,k=1}^n g^{ik}(x) A_i(x) f_k(x) \bigg) \sqrt{|g|} \:dx,
$$
where $f = \sum_{i=1}^n f_i(x) \,dx^i$.
In physics this is often referred to as the field "smeared out" with a test function $f$.
Of course the equation $dA=0$ is then equivalent to $A(\delta f)=0$ for all test functions $f \in \Omega^{p+1}_0(M)$.
Similarly, the wave equation~$\Box_p A=0$ is equivalent to $A(\Box_p f)=0$ for all $f \in \Omega^p_0(M)$.
When dealing with quantum fields, we shall always take this ``dual'' point of view.
Note that any cycle~$\gamma$ can be thought of as a co-closed distributional current. This means that
knowing $\int_\gamma A$ for all cycles is equivalent to knowing~$A(f)$ for all $f \in \Omega^1_0(M)$
with $\delta f=0$.

\section{The Field Algebras and the Gauge Ideals} \label{sec3}
In this section, we construct the field algebra of the quantized Maxwell field. Since we will be
dealing mostly with one-forms, we shall often omit the subscript~$p$ in the case $p=1$ and simply
write~$G_\pm$ for the advanced and retarded fundamental solutions and set~$G=G_+-G_-$.
The {\em{field algebra}}~$\mathcal{F}$ is defined to be the unital $*$-algebra generated by symbols
$A(f)$ for~$f \in \Omega^1_0(M)$ together with the relations
\begin{gather}
 f \mapsto A(f) \;\textrm{is linear},\\
 A(f) A(g) - A(g) A(f) = -i \,G(f,g),\\
 A(\square f) = 0,\quad \text{for all~$f \in  \Omega^1_0(M)$ with~$\delta f =0$}\:, \label{Arel} \\
 (A(f))^* = A(f).
\end{gather}
For every open subset $\mathcal{O} \subset M$, we define the
{\em{local field algebra}}~$\mathcal{F}(\mathcal{O}) \subset \mathcal{F}$
to be the sub-algebra generated by the $A(f)$ with $\supp(f) \subset \mathcal{O}$.
Inside $\mathcal{F}$, the {\em{algebra of observables}}~$\mathcal{A}$ is defined as the
unital subalgebra generated by $A(f)$ with $\delta f=0$. The local algebras of
observables~$\mathcal{A}(\mathcal{O})$ are given by~$\mathcal{A}(\mathcal{O}) = \mathcal{A}
\cap \mathcal{F}(\mathcal{O})$.

\begin{Remark} {\bf{(Choice of field algebra)}} {\em{
We point out that the relation~\eqref{Arel} is weaker than the more common
requirement $A(\delta d f)=0$ for all $f \in \Omega^1_0(M)$.
In particular, we do not impose that the field operators satisfy the homogeneous Maxwell equations.
This is a well-known feature of the Gupta-Bleuler formalism.
A general no-go theorem by Strocchi (see~\cite{strocchi0, strocchi, strocchi2}) states that under the
assumption of locality of the field, there is no non-trivial Lorentz covariant quantization of the
electromagnetic potential on four dimensional Minkowski space in which the homogeneous 
Maxwell equations are satisfied as an operator identity. As remarked in \cite[page~2199]{strocchi}, the
Gupta-Bleuler formalism circumvents this by relaxing the homogeneous Maxwell equation to a weak condition
on the physical subspace.
Our choice of field algebra together with the gauge ideal below (with~$\Lambda \equiv 0$
to obtain the Lorenz gauge) reflects the
construction in the original article~\cite{bleuleralt} as well as the presentation in
the standard textbooks on quantum electrodynamics.
}}  \QEDrem \end{Remark}

The physical interpretation of the algebra $\mathcal{A}(\mathcal{O})$ is that it consists of all the physical quantities that can be measured in the space-time region $\mathcal{O}$. In particular, if $g \in \Omega^2_0(\mathcal{O})$, then $A(\delta g)$ is an observable. Since $A(\delta g) = d A (g)$, this observable corresponds to the field
strength operator smeared out with the test function $g$. However, as explained at the end of the previous
section, the algebra of observables may also contain observables that correspond to 
smeared out measurements of $A$ along homologically non-trivial cycles. Thus it may be strictly larger than the algebra generated by $dA (g)$.

\begin{Def} 
 Let $\Lambda \in C^\infty(M)$. The {\bf{gauge ideal}}~$\mathcal{I}_\Lambda \subset \mathcal{F}$
 is the two-sided ideal generated by
 $$
   \left \{ A(\square f) - \Lambda(\delta \square f) \mid f \in \Omega^1_0(M) \right \}.
 $$
\end{Def} \noindent

\begin{Lemma} $\mathcal{A} \cap \mathcal{I}_\Lambda = \{0\}$.
\end{Lemma}
\Proof We first observe that
\beq \label{caprel}
K:= \square \ker \delta|_{\Omega^1_0(M)} \;=\;
\square \big( \Omega^1_0(M) \big) \cap \ker \delta|_{\Omega^1_0(M)} \:.
\eeq
Indeed, if~$f$ lies in $\square \ker \delta|_{\Omega^1_0(M)}$, then it is obviously both
in the image of~$\Box$ and co-closed. Conversely, if~$f$ lies in the intersection on the right,
then~$f=\square g$ and~$0 = \delta f$ and thus~$\square \delta g = 0$.
Since~$\delta g$ has compact support and solves the wave equation, it follows that~$\delta g=0$,
proving~\eqref{caprel}. We introduce the vector spaces
\[ W = \Omega^1_0(M) / K \qquad \text{and} \qquad
U = \square \big( \Omega^1_0(M) \big) / K\:,\quad
V = \ker \delta|_{\Omega^1_0(M)} / K\:. \]
Then
\beq \label{UcV}
U \cap V = \{ 0 \}\:.
\eeq
We introduce~$W^*_U$ and~$W^*_V$ as the subspaces of the dual space~$W^*$
consisting of elements that vanish on~$U$ and~$V$, respectively.

The algebra~${\mathcal{F}}$ has a natural filtration~${\mathcal{F}}_0 \subset
{\mathcal{F}}_1 \subset \cdots \subset {\mathcal{F}}$, where~${\mathcal{F}}_n$
is the span of products of the form~$A(f_1) \cdots A(f_m)$ with $m \leq n$.
Moreover, the linear map $\sigma_n: {\mathcal{F}}_n \rightarrow 
\bigotimes\nolimits_s^n W$ with kernel $\mathcal{F}_{n-1}$ mapping
$A(f_1) \cdots A(f_n)$ to $f_1 \otimes_s \cdots \otimes_s f_n$
is well-defined, where~$\otimes_s$ denotes the symmetric tensor product.

Let~${\mathfrak{f}} \in \mathcal{A} \cap \mathcal{I}_\Lambda$.
We want to show that~${\mathfrak{f}}$ vanishes. Thus assume by contradiction
that~${\mathfrak{f}} \neq 0$. Then there is a minimal~$n$
such that~${\mathfrak{f}} \in {\mathcal{F}}_n$ and~$F := \sigma_n({\mathfrak{f}}) \neq 0$.
Since~${\mathfrak{f}} \in {\mathcal{A}}$ is an observable, we know that~$F \in \otimes_s^n V$, and thus
\beq \label{c1}
\iota_\lambda F = 0 \qquad \text{for all~$\lambda \in W^*_V$} \:,
\eeq
where~$\iota_\lambda$ denotes the contraction with~$\lambda$.
On the other hand, as~${\mathfrak{f}} \in {\mathcal{I}}_\Lambda$,
we know that
\beq \label{c2}
(\lambda_1 \otimes_s \cdots \otimes_s \lambda_n) F = 0
\qquad \text{for all~$\lambda_i \in W^*_U$}\:.
\eeq
Combining~\eqref{c1} and~\eqref{c2}, we obtain by linearity that
\beq \label{c3}
(\lambda_1 \otimes_s \cdots \otimes_s \lambda_n) F = 0
\qquad \text{for all~$\lambda_i \in W^*_U + W^*_V$} \:.
\eeq
In view of~\eqref{UcV}, the set~$W^*_U+W^*_V$ is a point-separating subspace of~$W^*$.
Hence~\eqref{c3} implies that~$F=0$, a contradiction.
\QED
This lemma gives a canonical injective map $\mathcal{A} \to \mathcal{F}/\mathcal{I}_\Lambda$.

\begin{Remark} {\bf{(gauge transformations)}} {\em{
The analog of classical gauge transformations can be realized by the
algebra homomorphism
\beq \label{gaugetrans}
{\mathfrak{G}}_\Lambda \::\: A(f) \mapsto A(f) - (d \Lambda)(f) \:.
\eeq
This algebra homomorphism leaves the algebra of observables~${\mathcal{A}}$ invariant.
Moreover, it transforms the gauge ideals by
\beq \label{idtrans}
{\mathfrak{G}}_{\Lambda'} \mathcal{I}_\Lambda = \mathcal{I}_{\Lambda+\Lambda'} \:.
\eeq
Thus the gauge freedom is described in the algebraic formulation by the freedom in choosing
a gauge ideal.
}} \QEDrem \end{Remark}

In classical gauge theories, the time evolution is uniquely defined
only after a gauge-fixing procedure. In the same way, in our framework the 
time slice axiom holds only after dividing out the gauge ideal:
\begin{Prp} {\bf{(time slice axiom)}} Let~$U$ be an open neighborhood of a Cauchy surface in~$M$
and~$\Lambda \in C^\infty(M)$. Then
\[ {\mathcal{F}}(U)/\mathcal{I}_\Lambda = {\mathcal{F}}/\mathcal{I}_\Lambda \qquad \textrm{and} \qquad \mathcal{A}(U)=\mathcal{A}.\] 
\end{Prp}
\Proof Since the above gauge transformations leave~${\mathcal{F}}(U)$ invariant,
we can arrange in view of~\eqref{gaugetrans} and~\eqref{idtrans} that~$\Lambda=0$.
Let~$f \in \Omega^1_0(M)$. By Lemma~\ref{timeslicelemma} below there exist
forms~$h \in \Omega^1_0(M)$ and~$g \in \Omega^1_0(U)$ with~$\Box h = f-g$.
We conclude that~$A(f-g) = A(\Box h) \in {\mathcal{I}}_\Lambda$.
\QED

\begin{Lemma}\label{timeslicelemma}
 Suppose that~$U$ is an open neighborhood of a Cauchy surface~$\Sigma$ in~$M$. Then for
 every~ $f \in \Omega^\bullet_0(M)$ there exists $h \in \Omega^\bullet_0(M)$ such that the form
 \[f-\Box h\]
 is compactly supported in $U$. If~$df = 0$, then~$h$ can be chosen to be closed. If~$\delta f=0$,
 then~$h$ can be chosen to be co-closed.
\end{Lemma}
\Proof
 Let $\eta_+ \in C^\infty(M)$ and $\eta_-$ be non-negative smooth functions such that
 \begin{itemize}
  \item $\eta_+(x)^2 + \eta_-(x)^2 = 1$ for all $x \in M$.
  \item $\eta_+$ has future compact support and $\eta_-$ has past compact support.
  \item $\supp(d\eta_\pm) \subset U$.
 \end{itemize}
 Now one checks by direct computation that $h:= \eta_+^2 G_-(f) + \eta_-^2 G_+(f)$
 has the required properties. If $f$ is closed, we may take
 \[ h:= d \big( \eta_+ G_- (\eta_+ G_-(\delta f) ) +  \eta_- G_+ ( \eta_- G_+(\delta f)) \big).\]
Again one checks that $f-\Box h$ has compact support in $U$. Moreover, by construction,
$h$ is closed. A straightforward modification of this argument shows that~$h$ can be chosen to be co-closed
if $f$ is co-closed.
\QED

\section{Representations of~${\mathcal{F}}$} \label{sec4}
In Minkowski space, there is a unique vacuum state determined by
Poincar{\'e} invariance and the spectrum condition.
In general curved space-times, the lack of such a distinguished vacuum state
has led to alternative selection criteria for physical states.
The spectrum condition is then replaced by microlocal versions~\cite{radzikowski}.
Before introducing representations, we therefore recall some basic notions of microlocal analysis.

\subsection{Polarization Sets and Wavefront Sets of Bundle-Valued Distributions}
We denote by~$\PDO^m(M, E)$ the set of properly supported pseudo-differential operators
acting on sections of a vector bundle~$E \rightarrow M$.
More precisely, we work with polyhomogeneous symbols, i.e.\ symbols
in the H\"ormander classes~$S^m_\text{phg}$ defined in~\cite[Chapter~18]{hormanderIII}.
The principal symbol~$\sigma_A $ of a pseudo-differential operator~$A \in \PDO^m(M, E)$
is then a positive homogeneous section of degree~$m$ in~$C^\infty(\dot{T}^*M, \pi^* \End(E))$
(where~$\dot{T}^*M$ denotes the cotangent space with its zero section removed, and $\pi:\dot{T}^*M \to M$
is the canonical projection).
Following~\cite{dencker}, we define:
\begin{Def} Let~$u \in {\mathcal{D}}'(M; E)$ be a distribution with values in~$E$. Then
the {\bf{polarization set}}~$\WFpol(u)$ is defined by
\[ \WFpol(u) = \bigcap_{ \footnotesize{ \begin{matrix} A \in \PDO^0(M; E), \\
Au \in C^\infty(M; E) \end{matrix} } } {\mathcal{N}}_A \:, \]
where
\[ {\mathcal{N}}_A := \left\{
(x, \xi; v) \in \dot{T}^*M \times E 
\:\big|\: v \in E_x \text{ and } \sigma_A(x,\xi)\, v = 0 \right\} . \]
Moreover, the {\bf{wave front set}} can be defined by
\[ \WF(u) = \pi \Big( \WFpol(u) \setminus \dot{T}^*M \times \{0\} \Big) \:, \]
where~$\pi : \dot{T}^*M \times E \rightarrow \dot{T}^*M$ is the natural projection.
\end{Def}

\subsection{Gupta-Bleuler Representations} \label{secgenrep}
As is usual in the Gupta-Bleuler formalism,
the representation of the field algebra will not be
on a Hilbert space, but rather on a space equipped with an indefinite inner product.
Thus we let~$({\mathfrak{K}}, \la \cdot , \cdot \ra)$ be a locally convex topological vector spaced
endowed with an indefinite inner product. 

For a given real-valued function~$\Lambda \in C^\infty(M)$, we
let~$\pi$ be a representation of~${\mathcal{F}}$ on~${\mathfrak{K}}$
and~$\Omega \in {\mathfrak{K}}$ such that the following hold:
\begin{itemize}
\item[(a)] $\overline{\pi({\mathcal{F}})\, \Omega} = {\mathfrak{K}}$, \quad (cyclicity)
\item[(b)] $\pi({\mathcal{A}}) \,\Omega$ is a positive semi-definite
subspace~${\mathfrak{H}}_0 \subset {\mathfrak{K}}$ and~$\la \Omega, \Omega \ra=1$.
\item[(c)] $\pi({\mathcal{I}}_\Lambda)=0$,
\item[(d)] For any~$n \in \N$, the space~${\mathfrak{K}}_n$ defined by
\[ {\mathfrak{K}}_n = \overline{ \pi({\mathcal{F}}_n) \Omega } \subset {\mathfrak{K}} \]
is a Krein space (endowed with the inner product~$\la \cdot , \cdot \ra$ and the locally convex topology induced
by~${\mathfrak{K}}$). As before, ${\mathcal{F}}_n$
is the span of products of the form~$A(f_1) \cdots A(f_m)$ with $m\leq n$.
\item[(e)] microlocal spectrum condition:
\beq \label{npt}
\WF \Big( \pi( \underbrace{A(\cdot) \cdots A(\cdot)}_{\text{$m$~factors}}) \Omega \Big)
\subseteq \Gamma_m^+ \qquad \text{for all~$m$} \:,
\eeq
where~$\pi(A(\cdot) \cdots A(\cdot)) \Omega$ is a Krein-space-valued distribution.
The sets~$\Gamma_m^+$ are defined below.
\item[(f)] Gupta-Bleuler condition:
\[ \big\la \phi, \pi \big( A(df_1) - \Lambda(\Box f_1) \big) \cdots \pi \big( A(df_n) - \Lambda(\Box f_n) \big)\, \phi \big\ra = 0\]
for all~$\phi \in {\mathfrak{H}}_0$ and $n \geq 1$.
\end{itemize}
Then~$(\pi, {\mathfrak{K}}, \Omega)$ is called a {\bf{Gupta-Bleuler representation in the~$\Lambda$-gauge}}.
The distribution in~\eqref{npt} can be expressed in terms the~$n$-point distributions defined by
\[ \omega_n(f_1, \ldots, f_n) = \big\la \Omega, \pi \big( A(f_1) \cdots A(f_n) \big)\, \Omega \big\ra\:. \]
Note that~${\mathfrak{H}}_0$ is not a Hilbert space, because its inner product is only positive semi-definite.
Dividing out the null subspace
\[ {\mathcal{N}} = \{ \psi \in {\mathfrak{H}}_0 \:|\: \la \psi, \psi \ra = 0 \} \]
and forming the completion, one gets a Hilbert space,
which in the usual Gupta-Bleuler formalism is interpreted as the physical Hilbert space.
Note that as a consequence of the commutation relations the $A(df)$ generate a commutative $*$-algebra that commutes with
the observable algebra $\mathcal{A}$.

Here the sets~$\Gamma_m$ are defined as follows.
We denote the closed light cone and its boundary by
\begin{align*}
J^+ &= \left\{ (x, \xi) \:|\: g_x(\xi,\xi) \geq 0 \text{ and } \xi_0 \geq 0 \right\} \\
L^+ &= \left\{ (x, \xi) \:|\: g_x(\xi,\xi) = 0 \text{ and } \xi_0 \geq 0 \right\} .
\end{align*}

Let~$\mathcal{G}_k$ be the set of all finite graphs with vertices $\{ 1,\ldots,k\}$
such that for every element $G \in \mathcal{G}_k$ all edges occur in both
admissible directions. We write $s(e)$ and $r(e)$ for the source and the
target of an edge respectively. Following~\cite{brunetti+fredenhagen}, we define an
immersion of a graph~$G \in \mathcal{G}_k$ into the space-time~$M$ as an assignment
of the vertices $\nu$ of $G$ to points $x(\nu)$ in $M$, and of edges $e$
of $G$ to piecewise smooth curves $\gamma(e)$ in $M$ with source
$s(\gamma(e))=x(s(e))$ and range $r(\gamma(e))=x(r(e))$, together
with a covariantly constant causal co-vector field $\xi_e$ on $\gamma$ such that
\begin{enumerate}
 \item If $e^{-1}$ denotes the edge with opposite direction as $e$, then
   the corresponding curve $\gamma(e^{-1})$ is the inverse of $\gamma(e)$.
 \item For every edge $e$ the co-vector field $\xi_e$ is directed towards the
   future whenever $s(e) < r(e)$.
 \item $\xi_{e^{-1}}=-\xi_e$.
\end{enumerate}
We set
\begin{align*}
\Gamma_m := \Big\{
   (x_1,&\xi_1; \ldots; x_m,\xi_m) \in T^*M^m \backslash 0 \;\big|\;
   \textrm{there exists $G \in \mathcal{G}_m$}\\
&\textrm{and an immersion } (x,\gamma,\xi)
   \textrm{ of } G \textrm{ in } M \textrm{ such that } \\
&   x_i = x(i) \quad \text{for all}~i = 1,\ldots,m \text{ and }
\xi_i=-\sum_{e,\;s(e)=i} \xi_e(x_i) \Big\}\:.
\end{align*}
The set $\Gamma_m^+$ is defined as
$$
 \Gamma_m^+ = \Gamma_m + \left (J^+ \right)^m.
$$

The microlocal spectrum condition for $n$-point distributions
$$
 \mathrm{WF}(\omega_m) \subset \Gamma_m
$$
was introduced for scalar fields
by Bru\-netti, Fredenhagen and K\"ohler in~\cite{brunetti+fredenhagen},
who also showed that for quasi-free representations,
it suffices to verify their microlocal spectrum condition for the two-point functions (see also~\cite{sanders}).
Quasi-free representations are those which satisfy the Wick rule
\[ \omega_m(f_1,\ldots,f_m)=\sum_P \:\prod_r\: \omega_2(f_{(r,1)},f_{(r,2)}) \:, \]
where $P$ denotes a partition of the set $\{1,\ldots,m\}$ into subsets
which are pairings of points labeled by $r$.

The microlocal spectrum condition for quasi-free states of the Klein-Gordon field
was shown in~\cite{radzikowski} to be equivalent to the well-known Hadamard condition.
Moreover, the microlocal spectrum condition is a sufficient condition for the
construction of Wick polynomials (see~\cite{brunetti+fredenhagen, hollands2})
and interacting fields (see~\cite{brunetti3, hollands1}) in general globally hyperbolic space-times.
For this reason, the microlocal spectrum condition is
generally recognized to be a useful substitute for the spectrum
condition in Minkowski space valid in general globally hyperbolic space-times.

For scalar fields one can impose the microlocal spectrum condition for Hilbert space valued distributions
$$
 \mathrm{WF}(\underbrace{\phi(\cdot) \cdots \phi(\cdot)}_{\text{$m$~factors}} \Omega) \subset \Gamma_m^+.
$$
Note that if $(x_1,\xi_1;\ldots;x_n,\xi_n) \in \Gamma_m^+$ and $(x_n,-\xi_n;\ldots; x_1,-\xi_1) \in \Gamma_m^+$
then we have $(x_1,\xi_1;\ldots;x_n,\xi_n) \in \Gamma_m$.
Therefore the above microlocal spectrum condition implies the microlocal spectrum condition
for the corresponding $m$-point distribution. Following~\cite{brunetti+fredenhagen} one can show that the microlocal
spectrum condition for the Hilbert space valued distribution as satisfied for Wightman fields in Minkowski space-time
and also for quasifree Hadamard states of scalar fields in globally hyperbolic space-times.
For quasifree states of scalar fields the microlocal spectrum condition for states is therefore equivalent to the one
for Hilbert space valued distributions. This equivalence is based however on the Cauchy-Schwartz inequality 
which fails in Krein-spaces. We therefore require the microlocal spectrum condition to hold for the Krein space valued
distributions as above.
For Fock representations as described in Section \ref{sec43} the microlocal spectrum condition is equivalent to the requirement 
\[ \WF \Big( \pi \big( A( \cdot) \big) \,\Omega \Big) \subset J^+ \]
(where~$\pi(A(\cdot)) \,\Omega$ is again understood as a Krein-space-valued distribution), which is not difficult to show using the commutation relations
and Wick's theorem.

\begin{Remark} {\em{
Requiring the microlocal spectrum condition on the level of observables only
results in a slightly weaker condition that manifests itself in a condition on the polarization
set. Namely,  assume that~$u \in {\mathcal{D}}'(M, \Lambda^1(M))$ satisfies the condition
\beq \label{WFcon}
\WF(d u)  \subseteq \big\{ (x, \xi) \in \dot{T}^*M \:|\: \xi \in J^+ \big\} \:.
\eeq
Then
\beq \label{WFpolcon}
\WFpol(u) \subseteq \big\{ (x, \xi; v) \in \dot{T}^*M \times T^*M \:|\:
\xi \in J^+ \text{ or } \xi \sim v \big\} \:,
\eeq
where $\xi \sim v$ denotes linear dependence of $\xi$ and $v$.
Note that
\[ \WFpol(u) \subseteq \WFpol(du) + {\mathcal{N}}_d \]
with
\[ {\mathcal{N}}_d = \left\{ (x,\xi; v) \:\big|\: \sigma_d(x,\xi) \cdot v =0 \right\} = 
\Big\{ (x, \xi; v)  \:\big|\:
v \sim \xi \Big\} \:. \]
Microlocally, the set $\mathcal{N}_d$ corresponds to the so-called longitudinal photons.
Our condition (e) is stronger in that it imposes the microlocal spectrum condition in all directions, including those
corresponding to longitudinal photons.
We point out that the inverse implication \eqref{WFpolcon}$\Rightarrow$\eqref{WFcon} is in
general false, because the set~$\WFpol(u)$ only detects the highest order of the singularities.
}} \QEDrem \end{Remark}

\begin{Remark} (gauge invariance) {\em{
Note that property~(c) implements the field equation.
The property~(f), on the other hand, realizes the gauge condition~\eqref{gaugec},
but only if we take the inner product with a vector in~${\mathfrak{H}}_0$.
}} \QEDrem \end{Remark}

\subsection{Fock Representations} \label{sec43}
In order to construct Gupta-Bleuler-Fock representations of the field algebra $\mathcal{F}$,
one can proceed as follows. Let~$\kappa \::\: \Omega^1_0(M) \rightarrow {\mathscr{K}}$ be a real-linear
continuous map into a complex Krein space $\mathscr{K}$ with the following properties:
\begin{itemize}
\item[(i)] $\kappa(\Box f) = 0$ for all~$f \in \Omega^1_0(M)$
\item[(ii)] $\la \kappa(f), \kappa(f) \ra \geq 0$ if~$\delta f=0$.
\item[(iii)] $\im \la \kappa(f),\kappa(g) \ra = G(f,g)$
\item[(iv)] microlocal spectrum condition:
\[ \WF(\kappa)  \subseteq \big\{ (x, \xi) \in \dot{T}^*M \:|\: \xi \in J^+ \big\} \:. \]
\item[(v)] $\mathrm{span}_\mathbb{C} \mathrm{Rg}(\kappa)$ is dense in $\mathscr{K}$.
\item[(vi)] $\la \kappa(df), \kappa(g) \ra =0$ for all~$f \in C^\infty_0(M)$, $g \in \Omega^1_0(M)$
with either~$\delta g=0$ or with $g= dh$ for some $h \in C^\infty_0(M)$.
\end{itemize}

We introduce the Bosonic Fock space by
\beq \label{Kdef}
\mathfrak{K} =\bigoplus_{N=0}^\infty \hat{\bigotimes}_s^N \mathscr{K} \:,
\eeq
where~$\hat \otimes$ denotes the completed symmetric tensor products of Krein spaces.
Note that $\mathfrak{K}$ is an indefinite inner product space but does not have a canonical
completion to a Krein space.
For $\psi \in \mathscr{K}$, we let $a(\psi)$ be the annihilation operator and $a^*(\psi)$
be the creation operator, defined as usual by
\beq \label{adef}
\begin{split}
a^*(\psi) \,\phi_1 \otimes_s \ldots \otimes_s \phi_N &=  \psi \otimes_s 
\big(\phi_1 \otimes_s \ldots \otimes_s \phi_N \big) \\
a(\psi) \,\phi_1 \otimes_s \ldots \otimes_s \phi_N &= \la \psi, \phi_1 \ra \:\phi_2 \otimes_s \ldots \otimes_s
\phi_N \:.
\end{split}
\eeq
By construction, we have the canonical commutation relations
\beq \label{ccr}
\big[ a(\psi), a^*(\phi) \big] = \la \psi, \phi \ra \:.
\eeq

For each~$f \in \Omega^1_0(M)$ and a given $\Lambda \in C^\infty(M)$,
we let~$\hat{A}(f)$ be the following endomorphism of~${\mathfrak{K}}$:
\beq \label{hadef}
 \hat A(f) = \frac{1}{\sqrt{2}} \Big( a \big( \kappa(f) \big) + a^*\big( \kappa(f) \big) \Big) + d\Lambda(f) \:.
\eeq
Then the mapping
\[ \pi \::\: A(f) \mapsto \hat A(f) \]
extends to a $*$-representation $\pi$ of the field algebra $\mathcal{F}$ by operators
that are symmetric with respect to the indefinite inner product on  $\mathfrak{K}$.

\begin{Thm} \label{thm1}
The representation~$\pi$ is a Gupta-Bleuler representation.
\end{Thm}
\Proof
We need to check the properties (a)-(f) of a Gupta-Bleuler representation.
\begin{itemize}
\item[(a)] Cyclicity:
The Fock space is the direct sum of finite particle subspaces. Suppose that $N \geq 1$ and
let $P_N$ be the canonical projection onto the $N$-particle subspace
$\hat{\bigotimes}_s^N \mathscr{K}$. Since 
$$P_N \,\hat{A}(f_1) \cdots \hat{A}(f_N) \Omega
= \Big( \frac{1}{\sqrt{2}} \Big)^N \kappa(f_1) \otimes_s \cdots \otimes_s \kappa(f_N)$$
and the complex span of the range of $\kappa$ is dense in $\mathscr{K}$, we know that the complex span of
$\{ P_N \hat{A}(f_1) \cdots \hat{A}(f_N) \Omega \}$ is dense in $\hat{\bigotimes}_s^N \mathscr{K}$.
Therefore, if any element in $\bigoplus_{k=0}^{N-1} \hat{\bigotimes}_s^k \mathscr{K}$
can be approximated by elements in $\pi(\mathcal{F}) \Omega$, so can be any element in 
$\bigoplus_{k=0}^{N} \hat{\bigotimes}_s^k \mathscr{K}$. By induction in $N$, we conclude that
$\pi(\mathcal{F}) \Omega$ is dense in $\mathfrak{K}$. 
\item[(b)] is a direct consequence of (ii).
\item[(c)] follows from $\kappa(\Box(f))=0$ and the definition of $\hat A$ by direct computation.
\item[(d)] is clear by construction, because the finite tensor product of Krein spaces is a Krein space.
\item[(e)] The microlocal spectrum condition can be proved exactly as
in~\cite[Propositions~2.2 and~6.1]{strohmaier+verch} and~\cite[Proposition~4.3]{brunetti+fredenhagen}.
\item[(f)] The space~${\mathfrak{H}}_0$ is generated by vectors of the form
\[ \phi = \hat{A}(f_1) \cdots \hat{A}(f_n)\, \Omega \qquad \text{with~$\delta f_i = 0$} \:. \]
Using~\eqref{hadef}, we obtain
\[ \pi \big( A(df) - \Lambda(\Box f) \big)
= \frac{1}{\sqrt{2}} \Big( a \big( \kappa(df) \big) + a^*\big( \kappa(df) \big) \Big) \:. \]
As a consequence of the commutation relations~\eqref{ccr}
and the property~(vi), the operators~$a(\kappa(df))$ and~$a^*(\kappa(df))$ commute
with all the~$\hat{A}(f_k)$ and with each other. We conclude that~$a(\kappa(df)) \phi=0$ and
\begin{align*}
\big\la &\phi, \pi \big( A(df_1) - \Lambda(\Box f_1) \big) \ldots \pi \big( A(df_n) - \Lambda(\Box f_n) \big) \phi \ra \\
&= (\frac{1}{\sqrt{2}})^n \: \Big\la \phi, \Big( a \big( \kappa(df_1) \big) + a^*\big( \kappa(df_1) \big) \Big) 
\cdots \Big( a \big( \kappa(df_n) \big) + a^*\big( \kappa(df_n) \big) \Big)\, \phi \Big\ra = 0\:.
\end{align*}
\end{itemize}
\vspace*{-2.2em} \QED

\subsection{Generalized Fock Representations} \label{sec53}
In order to quantize the zero modes, we need to generalize the previous construction as follows.
Let~$\kappa \::\: \Omega^1_0(M) \rightarrow {\mathscr{K}}$ be a real-linear
continuous map into a complex Krein space $\mathscr{K}$ with the following properties:
\begin{itemize}
\item[(i)] $\kappa(\Box f) = 0$ for all~$f \in \Omega^1_0(M)$
\item[(ii)] $\la \kappa(f), \kappa(f) \ra \geq 0$ if~$\delta f=0$.
\item[(iii)] There is a bilinear form~$G_Z$ on~$\Omega^1_0(M) \times \Omega^1_0(M)$ with smooth integral kernel and the
following properties:
\begin{gather}
\im \la \kappa(f),\kappa(g) \ra + G_Z(f,g) = G(f,g) \label{GZ} \\
\text{The vector space~$Z:=\Omega^1_0(M) / \{f \,|\, G_Z(f,\cdot) = 0 \}$
is finite dimensional.} \label{Zdef}
\end{gather}
\item[(iv)] microlocal spectrum condition:
\[ \WF(\kappa)  \subseteq \big\{ (x, \xi) \in \dot{T}^*M \:|\: \xi \in J^+ \big\} \:. \]
\item[(v)] $\mathrm{span}_\mathbb{C} \mathrm{Rg}(\kappa)$ is dense in $\mathscr{K}$.
\item[(vi)] $\la \kappa(df), \kappa(g) \ra =0$ for all~$f \in C^\infty_0(M)$, $g \in \Omega^1_0(M)$
with either~$\delta g=0$ or with $g= dh$ for some $h \in C^\infty_0(M)$.
\end{itemize}
We introduce~${\mathfrak{K}}$ and~$a$ as in the previous section
(see~\eqref{Kdef} and~\eqref{adef}).
Let~$\nu \::\: \Omega^1_0(M) \rightarrow Z$ be the quotient map,
and~$\tilde{G}_Z$ the induced symplectic form on~$Z$.

We choose a complex structure~${\mathfrak{J}}$ on $Z$ such that~$K(\cdot,\cdot)
:=-\tilde{G}_Z(\cdot,{\mathfrak{J}} \cdot)$ is a real inner product. This complex structure then induces a canonical splitting 
$Z=Y \oplus \tilde Y$ into two $K$-orthogonal Lagrangian subspaces $Y$ and $\tilde Y$ such that the symplectic form is given
by $\tilde{G}_Z((x_1,x_2),(y_1,y_2)) = K(x_1,y_2)-K(x_2,y_1)$. Let $\mathrm{pr}_1$ and $\mathrm{pr}_2$ be the 
canonical projections and let $\nu_i := \mathrm{pr}_i \circ \nu$.
On the Schwartz space~$\mathcal{S}(Y, \C)$, we define~$\hat{A}_{\mathfrak{J}}(f) \in \End(\mathcal{S}(Y, \C))$ by
\[ \big( \hat{A}_{\mathfrak{J}}(f) \phi \big)(x) =   K \big( \nu_1(f), x \big)\: \phi(x)
+ i\, (D_{\mathfrak{J}\nu_2(f)} \phi)(x) \]
(where~$D_{\mathfrak{J}\nu_2(f)}$ denotes the derivative in the direction~${\mathfrak{J}\nu_2(f)} \in Y$).
A short computation using the identity
\begin{align*}
K &\big(\nu_1(f), x \big)\, \big(D_{\mathfrak{J}\nu_2(g)} \phi \big)(x) - D_{\mathfrak{J}\nu_2(g)}
\Big( K \big(\nu_1(f), x \big)\, \phi(x) \Big) \\
&= -\left( D_{\mathfrak{J}\nu_2(g)} K(\nu_1(f), x) \right) \phi(x) = -K \big( \nu_1(f), \nu_2 (g) \big)\: \phi(x)
\end{align*}
shows that~$\hat{A}_{\mathfrak{J}}$ satisfies the canonical commutation relations
\[ [\hat{A}_{\mathfrak{J}}(f), \hat{A}_{\mathfrak{J}}(g)] = -i G_Z(f,g)\:. \]

We now define~$\hat{A}(f)$ on~${\mathfrak{K}} \otimes {\mathcal{S}(Y, \C)}$ by
\[  \hat A(f) = \frac{1}{\sqrt{2}} \Big( a \big(\kappa (f) \big) + a^* \big( \kappa(f) \big) \Big)  \otimes \1
+\1 \otimes A_{\mathfrak{J}}(f) + d\Lambda(f) \:. \]

Since by assumption $G_Z$ has a smooth integral kernel, the wave front set of the distribution
$\1 \otimes A_{\mathfrak{J}}(f)$ is empty. A straightforward modification of the proof of Theorem~\ref{thm1}
leads to the following result.

\begin{Thm} Under the assumptions~{\rm{(i)--(vi)}} stated above,
the mapping~$\pi \::\: f \mapsto \hat{A}(f)$ defines a Gupta-Bleuler representation.
\end{Thm}

\section{Constructions for Ultrastatic Manifolds} \label{secultra}
In this section we assume that the manifold $M$ is ultrastatic, i.e.\ that it is of the
form~$M=\mathbb{R} \times \Sigma$ with metric of product type $g= dt^2 -h$, where
$h$ is a complete Riemannian metric on $\Sigma$. Then $M$ is globally hyperbolic and each 
$\Sigma_t:=\{t\} \times \Sigma$ is a Cauchy surface.
A one-form~$f \in \Omega^1(M)$ can be decomposed as
$$ f = f_0 \,dt + f_\Sigma \:, $$
where $f_0 \in C^\infty(M)$ and $f_\Sigma \in C^\infty(M) \otimes_\pi \Omega^1(\Sigma)$.
Here $\otimes_\pi$ denotes the projective tensor product of two locally convex spaces.
We can think of $f_\Sigma$ as a family of one-forms $f_\Sigma(t)$ on $\Sigma$ that depends
smoothly on the parameter $t$. Let
$$
 \Psi^f :=  \begin{pmatrix} f \\ \dot f \end{pmatrix},
$$
where $ \dot f:= \frac{df}{dt}$. The restriction $\Psi^f_t$ of $\Psi^f$ to the hypersurface
$\Sigma_t$ will be viewed as an element in $(C^\infty(\Sigma) \oplus \Omega^1(\Sigma))^2$.
We say that a one-form $f \in \Omega^1(M)$ has {\em{spatially compact support}} if
its restriction to any Cauchy surface has compact support. In particular, 
$\Psi^f_t \in (C^\infty_0(\Sigma) \oplus \Omega^1_0(\Sigma))^2$ for all $t \in \R$.
The set of spatially compact one-forms is denoted by $\Omega^1_{\mathrm{sc}}(M)$.

In view of the unique solvability of the Cauchy problem,
the set of smooth solutions~$\Omega^1_{\mathrm{sc}}(M) \cap \mathrm{ker}(\square)$ 
of the wave equation with spatially compact support can be identified with the space of initial
data with compact support on $\Sigma_0$. Thus, the map $f \mapsto \Psi^f_0$ defines an isomorphism
between $\Omega^1_{\mathrm{sc}}(M) \cap \mathrm{ker}(\square)$ and $(C^\infty_0(\Sigma) \oplus \Omega^1_0(\Sigma))^2$.
Since $G$ maps $\Omega^1_0(M)$ onto $\Omega^1_{\mathrm{sc}}(M) \cap \mathrm{ker}(\square)$,
the assignment
$$
 f \mapsto \Psi^{G(f)}_0
$$
defines a surjective map to the Cauchy data space $(C^\infty_0(\Sigma) \oplus \Omega^1_0(\Sigma))^2$.
There exists a natural symplectic form $\sigma$ on the Cauchy data space defined by
\beq \label{sigmadef}
\sigma \left( \begin{pmatrix} \f \\ \dot{\f} \end{pmatrix},
\begin{pmatrix} \g \\ \dot{\g} \end{pmatrix} \right)
= -\int_\Sigma \left( \f_0 \,\dot{\g}_0 - \dot{\f}_0 \,\g_0 \right) d\mu_\Sigma
+\int_\Sigma \left( \la \f_\Sigma, \dot{\g}_\Sigma \ra - \la \dot{\f}_\Sigma, \g_\Sigma \ra \right) d\mu_\Sigma\:,
\eeq
where $\la \cdot , \cdot \ra$ is the fibrewise inner product on forms on~$\Sigma$ induced by the Riemannian metric~$h$,
and~$\mu_\Sigma$ is the Riemannian measure on~$\Sigma$.
An elementary computation using Stokes' formula shows that (see for example~\cite[eq.~(4.6)]{baer+ginoux})
\[ G(f,g) = \sigma \big( \Psi^{Gf}, \Psi^{Gg} \big) \:. \]

\subsection{Absence of Zero Resonance States} \label{secreson}
Usually, the construction of ground states on ultrastatic space-times assumes the existence
of a spectral gap. In what follows, we shall generalize this construction significantly assuming
a weaker condition, which we now formulate. Let~$\Omega^p_{(2)}(\Sigma)$ be the space of
square-integrable $p$-forms on~$\Sigma$.
Since~$\Sigma$ is assumed to be complete, the Hodge Laplacian~$\Delta$
with domain of definition~$\Omega^p_0(\Sigma)$ is an essentially
self-adjoint operator on~$\Omega^p_{(2)}(\Sigma)$ .
We denote the self-adjoint extension again by~$\Delta$ with domain~${\mathscr{D}}(\Delta)$.
Let~$dE_z$ be the spectral measure of~$\Delta$.
Moreover, let~$\Omega^{p \perp}_{(2)}(\Sigma) \subset \Omega^p_{(2)}(\Sigma)$ be the orthogonal
complement of~$\ker \Delta$, and~$\Omega^{p \perp}_0(\Sigma) =\Omega^{p \perp}_{(2)}(\Sigma)
\cap \Omega^p_0(\Sigma)$.
Of course, $\Delta$ leaves~$\Omega^{p \perp}_{(2)}(\Sigma)$ invariant.
For simplicity, we denote its restriction to~$\Omega^{p \perp}_{(2)}(\Sigma)$ again by~$\Delta$.
Our constructions rely on the following condition:
\beq \label{A} \begin{split}
\text{(A)}\;\; & \text{The kernel of~$\Delta$ is finite dimensional,} \\[-0.5em]
& \text{and the
domain of the operator~$\Delta^{-\frac{1}{4}}$ contains~$\Omega^{p \perp}_0(\Sigma)$.}
\end{split}
\eeq
It is remarkable that for a large class of manifolds, this condition can be guaranteed under
topological conditions on the boundary at infinity. In fact, this condition is closely related to the absence of zero
resonance states. Namely, assume that the resolvent
family~$(\Delta-\lambda^2)^{-1}$ of the Laplacian on differential forms admits a meromorphic
continuation in the following sense. For suitably weighted $L^2$-spaces
\[ \H_{1} := L^2(\Sigma, \rho^{-1}\: d\mu_\Sigma)
\;\subset\; \Omega^\bullet_{(2)}(\Sigma) \;\subset\; \H_1^* = \H_{-1} := L^2(\Sigma, \rho\: d\mu_\Sigma) \]
with a positive weight function~$\rho \in C^\infty(\Sigma)$ that vanishes at infinity, 
we assume that the family of operators
\[ (\Delta-\lambda^2)^{-1} \::\: \H_{1} \rightarrow \H_{-1} \]
has a meromorphic extension to a neighborhood of~$\lambda=0$,
with the property that the negative Laurent coefficient are operators of finite rank.
This assumption is well-known to be satisfied for odd-dimensional manifolds which are
isometric to~$\R^{2n+1}$ outside compact sets (see for example~\cite{melrose}).
Moreover, meromorphic continuations have been established for manifolds with cylindrical
ends~\cite{melrose2} in the context of the Atiyah-Patodi-Singer index theorem.
It follows from standard glueing constructions and the meromorphic Fredholm theorem that
the meromorphic properties of the resolvent are stable under compactly supported metric
and topological perturbations and therefore only depend on the structure near infinity.

Under these assumptions, there exist finite-rank operators~$A, B : \H_{1} \rightarrow \H_{-1}$
such that for any~$f \in \Omega^\bullet_0$, the measure~$d\la f, E_z f \ra$ has the representation
\beq \label{formula}
d\la f, E_z f \ra = \bigg( \la f, A f \ra\, \delta(z) + \la f, B f\ra\: \frac{\Theta(z)}{\sqrt{z}} 
+ \big\la f, C(\sqrt{z}) f \big\ra\:\Theta(z)  \bigg) \,dz \:,
\eeq
where~$\Theta$ denotes the Heaviside step function, and~$C$ is a holomorphic family
of operators with values in the bounded operators~$\L(\H_{1}, \H_{-1})$.
The operator~$A$ is in fact the orthogonal projection onto~$\ker \Delta$.
Since~\eqref{formula} remains true for vectors in~$\ker \Delta$ if we set~$B=0$, $C=0$ and~$A=1$,
we can extend this formula by linearity to the domain~$\Omega^{\bullet}_0(\Sigma) + \ker(\Delta)$.
Using the spectral theorem, one easily sees that the above condition~(A) is equivalent to the vanishing
of the operator~$B$ on the subspace $\Omega^{\bullet \perp}_0(\Sigma)$.

Vectors in the range of~$B$ that are not in $\ker \Delta$ are commonly referred to as {\em{zero resonant states}}.
The topological significance of these states was first pointed out by Atiyah, Patodi and
Singer~\cite{atiyah+patodi+singer} and elaborated in~\cite{mueller, melrose2}
in the case of manifolds with cylindrical ends (see also the introduction in~\cite{mueller+strohmaier}).
For another class of manifolds, referred to as non-parabolic at infinity, it was pointed out
by Carron~\cite{carron2} that the existence of certain non-square-integrable harmonic forms
depends only on the geometry near infinity, the obstruction being an index of a certain Dirac operator.
In many situations, it can be shown that these non-square-integrable harmonic forms correspond to
zero resonant states~\cite{carron, wang}.

In order to illustrate that assumption~(A) is stable and holds for a large class of manifolds,
we now work out the above connections in the case of odd-dimensional manifolds which are
isometric to~$\R^{2n+1}$ outside compact sets.
This covers the physically interesting case of three space dimensions.
Our results could be extended to even dimensions by a straightforward analysis of
the logarithmic terms that are known to be present in the corresponding expansion~\eqref{formula}.

\begin{Prp} Let~$(\Sigma^{2n+1}, g)$ with~$n \geq 1$ be a complete Riemannian manifold 
which is  Euclidean at infinity in the sense that there
exist compact subsets~$K_1 \subset \Sigma$ and~$K_2 \subset \R^{2n+1}$ such
that~$\Sigma \setminus K_1$ is isometric to~$\R^{2n+1} \setminus K_2$.
Then the operator~$B$ in~\eqref{formula} vanishes on $\Omega^{p \perp}_0(\Sigma)$,
and condition~(A) in~\eqref{A} is satisfied.
\end{Prp}
\Proof By the above, it suffices to show that~$B|_{\Omega^{p \perp}_0(\Sigma)}=0$.
Following~\cite{carron2}, we introduce the Sobolev space~$W(\Lambda^\bullet T^*\Sigma)$ as the completion
of~$\Omega^\bullet_0(\Sigma)$ with respect to the quadratic form
\beq \label{form}
\int_U |\alpha|^2 \:d\mu_\Sigma + \int_\Sigma \Big( |d\alpha|^2 + |\delta \alpha|^2 \Big)\, d\mu_\Sigma\:,
\eeq
where~$U$ is a non-empty relatively compact open subset of~$\Sigma$.
Note that for~$\Sigma = \R^{2n+1}$, the
space~$\{ \alpha \in W(\Lambda^\bullet T^*\Sigma) \mid d \alpha +\delta \alpha = 0 \}$ is zero
provided that~$n \geq 1$.
As shown in~\cite[Theorem~0.6]{carron2}, the number
\[ \dim \frac{\{ \alpha \in W(\Lambda^\bullet T^*\Sigma) \mid d \alpha +\delta \alpha = 0 \}}{\ker(\Delta)} \]
depends only on the geometry of~$\Sigma$ near infinity. Therefore,
it is enough to show that the range of~$B$ is contained in~$W(\Lambda^\bullet T^*\Sigma)$.

To this end, we must show that for every zero-resonance state~$u \in \text{rg}\,B$,
there is a sequence~$u_n \in \Omega^\bullet_0(\Sigma)$ which converges to~$u$
in~$W(\Lambda^\bullet T^*\Sigma)$.
If~$\chi \in C^\infty_0(\R)$ is an even real-valued function with~$\int_\R \chi(x) \,dx=0$,
then its Fourier transform~$\hat{\chi}(\lambda) \in {\mathcal{S}}(\R)$ is a Schwartz function
that vanishes at~$\lambda=0$. We choose the function~$\chi$ with the additional property that
\[ \frac{1}{\sqrt{2 \pi}} \int_0^\infty \frac{\hat{\chi}\big(\sqrt{z} \big)}{\sqrt{z}}\: dz = 1 \:. \]
Moreover, for any~$\varepsilon>0$ we define
\[ \chi_\varepsilon(x) = \chi(\varepsilon x) \qquad \text{and thus} \qquad
\hat{\chi}_\varepsilon(\lambda) = \frac{1}{\varepsilon}\: \hat{\chi} \Big( \frac{\lambda}{\varepsilon} \Big) . \]
Since~$B$ has finite rank, there exists a compactly supported section~$v \in \Omega^\bullet_0(\Sigma)$
such that~$u = Bv$. By finite propagation speed of the operator~$\cos(t \Delta^\frac{1}{2})$, 
the section $u_\varepsilon := \hat{\chi}_\varepsilon(\Delta^{\frac{1}{2}})(v)$ is again compactly supported. We want to show that
\[ \lim_{\varepsilon \searrow 0} u_\varepsilon = u \qquad \text{with convergence
in~$W(\Lambda^\bullet T^*\Sigma)$}\:. \]
First, it follows from~\eqref{formula} that~$u_\varepsilon$ converges in~$\H_{-1}$ to $u$.
This implies in particular that
\[ \lim_{\varepsilon \searrow 0} \int_U |u_\varepsilon - u|^2\, d\mu_\Sigma = 0 \]
(where~$U$ is again the relatively compact set in~\eqref{form}). Next, again using~\eqref{formula},
\[ \int_\Sigma \left( |d u_\varepsilon|^2 + |\delta u_\varepsilon|^2 \right) d\mu_\Sigma = \la u_\varepsilon, \Delta u_\varepsilon \ra =\int_0^\infty z \,|\hat{\chi}_\varepsilon(\sqrt{z})|^2\: d \la v, E_z v \ra
\xrightarrow{\varepsilon \searrow 0} 0 \:, \]
showing that~$u_\varepsilon$ converges in~$W$. Since it converges in~$\H_{-1}$ to~$u$,
the limit in~$W$ is again~$u$.
\QED

\subsection{Construction of~$\kappa$} \label{seckappa}
We assume throughout this section that condition~(A) in~\eqref{A} holds.
We choose the Krein space~${\mathscr{K}}$ as
\[ {\mathscr{K}} = \big(-\Omega^{0 \perp}_{(2)}(\Sigma) \oplus 
\Omega^{1 \perp}_{(2)}(\Sigma) \big) \otimes_\R \C \:. \]
Our assumptions imply that the operator~$\Delta^{s}$ has a trivial kernel on~$\Omega^{p \perp}_{(2)}(\Sigma)$
and is densely defined for all~$s \in \R$. We introduce the spaces
\[ \H^s = \overline{{\mathscr{D}} \big( \Delta^\frac{s}{2} \big)\: }^{\|.\|_s} \:, \]
where the bar denotes the completion with respect to the norm~$\|\phi\|_s := \|\Delta^\frac{s}{2} \phi\|$
of the subspaces~${\mathscr{D}} (\Delta^\frac{s}{2}) \subset \Omega^{p \perp}_{(2)}(\Sigma)$.
It follows from the spectral calculus that~$\H^s$ is the topological dual of~$\H^{-s}$. Moreover, it
is obvious that
\[ {\mathscr{D}}(\Delta^s) \subset \H^{2s} \:, \]
with continuous inclusion. Next, the following map is continuous:
 \[ \Delta^t \::\: \H^s \rightarrow \H^{s-2t} \qquad \text{for all~$t \in \R$}\:. \]
Furthermore, using that~$\Delta$ commutes with all projections
onto the form degree and~$\Delta = (d_\Sigma + \delta_\Sigma)^2$, we also have the
continuous mappings
\[ d_\Sigma, \delta_\Sigma \::\: \H^s \rightarrow \H^{s-1}\:, \]
which commute with~$\Delta^t$ in the sense that
\[ \Delta^t \,d_\Sigma = d_\Sigma \,\Delta^t \qquad \text{and} \qquad
\Delta^t \,\delta_\Sigma = \delta_\Sigma \,\Delta^t \]
as continuous operators from~${\mathscr{D}}\big( \Delta^s \big)$ to~${\mathscr{D}}\big( \Delta^{s-2t-1} \big)$.
Finally, the adjoints with respect to the dual pairings~$\H^s$ and~$\H^{-s}$ are computed as usual, i.e.\
\[ (\Delta^t)^* = \Delta^t\:,\qquad d_\Sigma^* = \delta_\Sigma\:,\qquad
\delta_\Sigma^* = d_\Sigma \:. \]

In the following computations, by~$\la \cdot , \cdot \ra$ we denote the dual pairing
between the spaces~$\H^s$ and~$\H^{-s}$. We define the maps~$\tau$ and~$\kappa$ by
\begin{align}
\tau \,:\, \big( \Omega^{0 \perp}_0(\Sigma) \oplus \Omega^{1 \perp}_0(\Sigma) \big)^2
&\rightarrow {\mathscr{K}} \:, \nonumber \\
\begin{pmatrix} \f \\ \dot{\f} \end{pmatrix} &\mapsto
\left( \Delta^\frac{1}{4} \f_0 + i \Delta^{-\frac{1}{4}} \dot{\f}_0 \right) \oplus
\left( \Delta^\frac{1}{4} \f_\Sigma + i \Delta^{-\frac{1}{4}} \dot{\f}_\Sigma \right) \label{taudef} \\
\kappa \::\: \Omega^1_0(M) \rightarrow {\mathscr{K}}\:,\quad f &\mapsto \tau(P_\perp \Psi_0^{Gf})\:,
\end{align}
where~$P_\perp$ is a projection onto~$\Omega^{\bullet \perp}_0(\Sigma)$ such that $1-P_\perp$ has a
smooth integral kernel. In case $B=0$
this projection can be chosen to be the orthogonal projection, as then $P_\perp \Omega^\bullet_0(\Sigma)$ is in the domain of $\Delta^{-\frac{1}{4}}$.
\begin{Prp} \label{prp52}
The mapping~$\kappa$ has the following properties:
\begin{itemize}
\item[(i)] $\kappa(\Box f) = 0$ for all~$f \in \Omega^1_0(M)$.
\item[(ii)] $\la \kappa(f), \kappa(f) \ra \geq 0$ if~$\delta f=0$.
\item[(iii)] $\im \big( \big\la \kappa f,\kappa g) \big\ra \big) = G(f,g)$ for all~$f,g \in \Omega^1_0(M)$
with~$\Psi^{Gf}_0 \perp \ker \Delta$.
\item[(v)] $\mathrm{span}_\mathbb{C} \mathrm{Rg}(\kappa)$ is dense in $\mathscr{K}$.
\item[(vi)] $\la \kappa(df), \kappa(g) \ra =0$ for all~$f \in C^\infty_0(M)$, $g \in \Omega^1_0(M)$
with either~$\delta g=0$ or with $g= dh$ for some $h \in C^\infty_0(M)$.
\end{itemize}
\end{Prp}
\Proof The properties~(i) and~(v) hold by construction.
For both one-forms and functions we set
\[ \begin{pmatrix} \f \\ \dot{\f} \end{pmatrix}(t) = P_\perp \Psi_t^{Gf} \:, \]
and similarly for~$g$. 
Moreover, we denote the negative of the $L^2$ inner product on $\Sigma_0$ 
by $\la \cdot, \cdot \ra_\Sigma$, i.e.
$\la \f, \g \ra_\Sigma$ is the $L^2$-inner product of the restrictions of $\f$ and $\g$ to
to $\Sigma_0$. The convention is chosen so that this inner product coincides with the Krein space inner product and the notation
makes it clear that integration is over $\Sigma_0$ and not over the entire space $M$.
Then the computation
\begin{align*}
\im &\bigg\la \tau \bigg( \!\!\begin{pmatrix} \f \\ \dot{\f} \end{pmatrix}\!\! \bigg),
\tau \bigg( \!\!\begin{pmatrix} \g \\ \dot{\g} \end{pmatrix}\!\! \bigg) \bigg\ra \\
&= -\im \Big\la \Delta^\frac{1}{4} \f_0 + i \Delta^{-\frac{1}{4}} \dot{\f}_0, \,
\Delta^\frac{1}{4} \g_0 + i \Delta^{-\frac{1}{4}} \dot{\g}_0 \Big\ra_{\Sigma} \\
&\quad\, +\im \Big\la \Delta^\frac{1}{4} \f_\Sigma + i \Delta^{-\frac{1}{4}} \dot{\f}_\Sigma, \,
\Delta^\frac{1}{4} \g_\Sigma + i \Delta^{-\frac{1}{4}} \dot{\g}_\Sigma \Big\ra_{\Sigma} \\
&= -\big\la \Delta^\frac{1}{4} \f_0, \Delta^{-\frac{1}{4}} \dot{\g}_0 \big\ra_{\Sigma} 
+\big\la \Delta^{-\frac{1}{4}} \dot{\f}_0, \Delta^\frac{1}{4} \g_0 \big\ra_{\Sigma} \\
&\quad\, +\big\la \Delta^\frac{1}{4} \f_\Sigma, \Delta^{-\frac{1}{4}} \dot{\g}_\Sigma \big\ra_{\Sigma} 
-\big\la \Delta^{-\frac{1}{4}} \dot{\f}_\Sigma, \Delta^\frac{1}{4} \g_\Sigma \big\ra_{\Sigma} \\
&=-\big\la \f_0, \dot{\g}_0 \big\ra_{\Sigma} +\big\la \dot{\f}_0, \g_0 \big\ra_{\Sigma} 
+\big\la \f_\Sigma, \dot{\g}_\Sigma \big\ra_{\Sigma} -\big\la \dot{\f}_\Sigma, \g_\Sigma \big\ra_{\Sigma}
=\sigma \bigg(\!\! \begin{pmatrix} \f \\ \dot{\f} \end{pmatrix} \!
,\!\begin{pmatrix} \g \\ \dot{\g} \end{pmatrix} \!\!\bigg)
\end{align*}
together with~\eqref{sigmadef} yields~(iii). To prove~(ii), we first note that
\begin{align}
\bigg\la &\tau \bigg( \!\!\begin{pmatrix} \f \\ \dot{\f} \end{pmatrix}\!\! \bigg),
\tau \bigg( \!\!\begin{pmatrix} \f \\ \dot{\f} \end{pmatrix}\!\! \bigg) \bigg\ra \nonumber  
 =-\big\la \Delta^{\frac{1}{2}} \f_0, \f_0 \big\ra_{\Sigma} - \big\la \Delta^{-\frac{1}{2}} \dot{\f}_0, \dot{\f}_0 \big\ra_{\Sigma}\\
 & +\big\la \Delta^{\frac{1}{2}} \f_\Sigma, \f_\Sigma \big\ra_{\Sigma} + \big\la \Delta^{-\frac{1}{2}}
\dot{\f}_\Sigma, \dot{\f}_\Sigma \big\ra_{\Sigma}\:. \label{inter}
\end{align}
Since~$G f$ solves the wave equation,
\begin{align}
0 &= \Box \f = (-\ddot{\f}_0 - \Delta \f_0) \,dt + (-\ddot{\f}_\Sigma - \Delta \f_\Sigma) \label{sol} \\
\intertext{Moreover, if~$\delta f=0$, we have}
0 &= \delta \f = -\dot{\f}_0 - \delta_\Sigma \f_\Sigma \label{del0} \:.
\end{align}
The last equality in~\eqref{del0} can be verified with the computation
\begin{align*}
\la \delta \f, \varphi \ra &= \la f, d \varphi \ra =
\la \f_0 \,dt + \f_\Sigma, \dot{\varphi}\, dt + d_\Sigma \varphi \ra \\
&= \la \f_0, \dot{\varphi} \ra - \la \f_\Sigma, d_\Sigma \varphi \ra
\overset{(*)}{=} - \la \dot{\f}_0, \varphi \ra - \la \delta_\Sigma \f_\Sigma, \varphi \ra
= -\la (\dot{\f}_0 + \delta_\Sigma \f_\Sigma), \varphi \ra\:,
\end{align*}
where in~$(*)$ we integrated by parts. Then
\begin{align*}
-\big\la &\Delta^{-\frac{1}{2}} \dot{\f}_0, \dot{\f}_0 \big\ra_{\Sigma} +\big\la \Delta^{\frac{1}{2}} \f_\Sigma, \f_\Sigma \big\ra_{\Sigma} \\
&= -\big\la \Delta^{-\frac{1}{2}} (d_\Sigma \delta_\Sigma) \f_\Sigma, \f_\Sigma \big\ra_{\Sigma}
+\big\la \Delta^{-\frac{1}{2}} (d_\Sigma \delta_\Sigma + \delta_\Sigma d_\Sigma) \f_\Sigma, \f_\Sigma \big\ra_{\Sigma} \\
&= \big\la \Delta^{-\frac{1}{2}} \:\delta_\Sigma d_\Sigma \f_\Sigma, \f_\Sigma \big\ra_{\Sigma}
= \big\la \Delta^{-\frac{1}{2}} \:d_\Sigma \f_\Sigma, d_\Sigma \f_\Sigma \big\ra_{\Sigma} \geq 0\:.
\end{align*}
Moreover, differentiating~\eqref{del0} with respect to~$t$ and using~\eqref{sol} gives
\[ \delta_\Sigma \dot{\f}_\Sigma = \Delta \f_0 \:. \]
Hence
\begin{align*}
-\big\la &\Delta^{\frac{1}{2}} \f_0, \f_0 \big\ra_{\Sigma} 
+ \big\la \Delta^{-\frac{1}{2}} \dot{\f}_\Sigma, \dot{\f}_\Sigma \big\ra_{\Sigma} \\
&=-\big\la \Delta^{-\frac{3}{2}} \Delta \f_0, \Delta \f_0 \big\ra_{\Sigma}
+ \big\la \Delta^{-\frac{3}{2}} \Delta \dot{\f}_\Sigma, \dot{\f}_\Sigma \big\ra_{\Sigma} \\
&=-\big\la \Delta^{-\frac{3}{2}} \delta_\Sigma \dot{\f}_\Sigma, \delta_\Sigma \dot{\f}_\Sigma \big\ra_{\Sigma}
+ \big\la \Delta^{-\frac{3}{2}} (d_\Sigma \delta_\Sigma + \delta_\Sigma d_\Sigma)
\dot{\f}_\Sigma, \dot{\f}_\Sigma \big\ra_{\Sigma} \\
&= \big\la \Delta^{-\frac{3}{2}} \delta_\Sigma d_\Sigma \dot{\f}_\Sigma, \dot{\f}_\Sigma \big\ra_{\Sigma}
= \big\la \Delta^{-\frac{3}{2}} d_\Sigma \dot{\f}_\Sigma, d_\Sigma \dot{\f}_\Sigma \big\ra_{\Sigma} \geq 0\:.
\end{align*}
This shows~(ii).

In order to prove~(vi) in case $\delta g =0$, we polarize~\eqref{inter} to obtain
\begin{align*}
\la \kappa(d f), \kappa(g) \ra 
&=-\big\la \Delta^{\frac{1}{2}} \dot{\f}, \g_0 \big\ra_{\Sigma} - \big\la \Delta^{-\frac{1}{2}} \ddot{\f}, \dot{\g}_0 \big\ra_{\Sigma}
+\big\la \Delta^{\frac{1}{2}} d_\Sigma \f, \g_\Sigma \big\ra_{\Sigma} + \big\la \Delta^{-\frac{1}{2}}
d_\Sigma \dot{\f}, \dot{\g}_\Sigma \big\ra_{\Sigma} \\
&=-\big\la \Delta^{\frac{1}{2}} \dot{\f}, \g_0 \big\ra_{\Sigma} + \big\la \Delta^{\frac{1}{2}} \f, \dot{\g}_0 \big\ra_{\Sigma}
+\big\la \Delta^{\frac{1}{2}} d_\Sigma \f, \g_\Sigma \big\ra_{\Sigma} + \big\la \Delta^{-\frac{1}{2}}
d_\Sigma \dot{\f}, \dot{\g}_\Sigma \big\ra_{\Sigma} \\
&=-\big\la \Delta^{\frac{1}{2}} \dot{\f}, \g_0 \big\ra_{\Sigma} - \big\la \Delta^{\frac{1}{2}} \f,
\delta_\Sigma \g_\Sigma \big\ra_{\Sigma}
+\big\la \Delta^{\frac{1}{2}} d_\Sigma \f, \g_\Sigma \big\ra_{\Sigma} + \big\la \Delta^{-\frac{1}{2}}
d_\Sigma \dot{\f}, \dot{\g}_\Sigma \big\ra_{\Sigma} \\
&= 0 \:,
\end{align*}
where we have used the wave equations~$\Box \f =0$ and~$\Box \g=0$
together with the identity~$0 = \delta \g = -\dot{\g}_0- \delta_\Sigma \g_\Sigma$.

Similarly, if $g \in C^\infty_0(M)$ we have
\begin{align*}
\la \kappa(d f), \kappa(d g) \ra & = 
-\big\la \Delta^{\frac{1}{2}} \dot{\f}, \dot \g \big\ra_\Sigma - \big\la \Delta^{-\frac{1}{2}} \ddot{\f}, \ddot{\g} \big\ra_\Sigma
+\big\la \Delta^{\frac{1}{2}} d_\Sigma \f, d_\Sigma \g \big\ra_\Sigma + \big\la \Delta^{-\frac{1}{2}}
d_\Sigma \dot{\f}, d_\Sigma \dot{\g} \big\ra_\Sigma \\
&= -\big\la \Delta^{\frac{1}{2}} \dot{\f}, \dot \g \big\ra_\Sigma - \big\la \Delta^{\frac{3}{2}} \f, \g \big\ra_\Sigma
+\big\la \Delta^{\frac{1}{2}} d_\Sigma \f, d_\Sigma \g \big\ra_\Sigma + \big\la \Delta^{-\frac{1}{2}}
d_\Sigma \dot{\f}, d_\Sigma \dot{\g} \big\ra_\Sigma \\
&=-\big\la \Delta^{\frac{1}{2}} \dot{\f}, \dot \g \big\ra_\Sigma - \big\la \Delta^{\frac{3}{2}} \f, \g \big\ra_\Sigma
+\big\la \Delta^{\frac{3}{2}}  \f,  \g \big\ra_\Sigma + \big\la \Delta^{\frac{1}{2}}
 \dot{\f},  \dot{\g} \big\ra_\Sigma = 0 \:.\\
\end{align*}
This concludes the proof.
\QED

In view of~\eqref{GZ}, we define
\beq \label{GZform}
G_Z(f,g) = G(f,g) - \im \la \kappa(f),\kappa(g) \ra \:.
\eeq
Proposition~\ref{prp52}~(iii) has the following implication.
\begin{Corollary} \label{cor53} The symplectic vector space~$Z$ defined by~\eqref{Zdef} is canonically isomorphic
to~$\left( \ker \Delta \cap \left(\Omega^0_{(2)}(\Sigma) \oplus \Omega^1_{(2)}(\Sigma) \right) \right)^2$ with 
the restriction of the standard symplectic form defined in (\ref{sigmadef}). With the complex structure
\[
 {\mathfrak{J}} \left( \begin{pmatrix} (\f_0, \f_\Sigma) \\ (\g_0, \g_\Sigma) \end{pmatrix} \right)=  \left( \begin{pmatrix} (\g_0, -\g_\Sigma) \\ (-\f_0, \f_\Sigma) \end{pmatrix} \right)
\]
the associated inner product is the usual $L^2$-inner product.
\end{Corollary}

Summarizing the results of this section, we come to the following conclusion.
\begin{Thm} Let~$(\Sigma, g)$ be a Riemannian manifold satisfying assumption~{\rm{(A)}}
in~\eqref{A}. The above mapping~$\kappa$ together with the form~$G_Z$
given by~\eqref{GZform} and the complex structure ${\mathfrak{J}}$ given above defines a generalized Fock representation (see Section~\ref{sec53})
in the ultrastatic space-time~$(\R \times \Sigma, dt^2-g)$,
thereby giving rise to a Gupta-Bleuler representation of~${\mathcal{F}}$ (see Section~\ref{secgenrep}).
\end{Thm}

\subsection{Positivity of the Energy and the Microlocal Spectrum Condition}
It is obvious from Definition~\ref{taudef} that~$\tau$ is injective and that its image is dense in~$\K$.
We introduce the operator~$H$ on~${\mathfrak{K}}$ by
\beq \label{Hdef}
H = \begin{pmatrix} \Delta^{\frac{1}{2}} & 0 \\ 0 & \Delta^{\frac{1}{2}} \end{pmatrix}
\eeq
(acting in the first component on functions and on the second component on one-forms).
Then
\begin{align*}
H\, \tau \begin{pmatrix} \f \\ \dot{\f} \end{pmatrix}
&= \left( \Delta^{-\frac{1}{4}} \Delta \f_0 + i \Delta^{\frac{1}{4}} \dot{\f}_0 \right) \oplus
\left( \Delta^{-\frac{1}{4}} \Delta \f_\Sigma + i \Delta^{\frac{1}{4}} \dot{\f}_\Sigma \right) \\
&= i \left( \Delta^{\frac{1}{4}} \dot{\f}_0 + i \Delta^{-\frac{1}{4}} (-\Delta) \f_0 \right) \oplus
\left( \Delta^{\frac{1}{4}} \dot{\f}_\Sigma + i \Delta^{-\frac{1}{4}} (-\Delta) \f_\Sigma \right) \\
&= \rmi \tau \begin{pmatrix} 0 & 1 \\ -\Delta & 0 \end{pmatrix}
\begin{pmatrix} \f \\ \dot{\f} \end{pmatrix} 
\overset{\eqref{sol}}{=} \rmi \partial_t \,\tau \begin{pmatrix} \f \\ \dot{\f} \end{pmatrix} \:.
\end{align*}
Assume that~$\delta f=0$. Then, similar as above,
\begin{align*}
\lefteqn{ \bigg\la \tau \bigg( \!\!\begin{pmatrix} \f \\ \dot{\f} \end{pmatrix}\!\! \bigg), H \,
\tau \bigg( \!\!\begin{pmatrix} \f \\ \dot{\f} \end{pmatrix}\!\! \bigg) \bigg\ra } \\
&=-\big\la \Delta \f_0, \f_0 \big\ra_\Sigma - \big\la \dot{\f}_0, \dot{\f}_0 \big\ra_\Sigma
+\big\la \Delta \f_\Sigma, \f_\Sigma \big\ra_\Sigma + \big\la
\dot{\f}_\Sigma, \dot{\f}_\Sigma \big\ra_\Sigma \\
&= \big\la d_\Sigma \f_\Sigma, d_\Sigma \f_\Sigma \big\ra_\Sigma
+ \big\la \Delta^{-1} d_\Sigma \dot{\f}_\Sigma, d_\Sigma \dot{\f}_\Sigma \big\ra_\Sigma \geq 0\:.
\end{align*}

\begin{Prp}\label{prop5.5}
The mapping~$\kappa$ satisfies the
\begin{itemize}
\item[(iv)] microlocal spectrum condition:
\[ \WF(\kappa( \cdot )) \subset  J^+ \]
\end{itemize}
\end{Prp}
\Proof Let~$T_t$ be the operator that shifts functions and distributions by~$t \in \R$ in time.
By construction of~$H$, we know that, modulo smooth functions,
\[ \kappa( T_{-t} f ) = e^{i H t}\, \kappa(f) \: \mathrm{mod}\, C^\infty .\]
For a given point~$x_0 \in \Sigma$, we choose a chart~$x$ and a bundle chart.
Let~$\chi \in C^\infty_0(\Sigma;T^*M)$ be any smooth section supported in our chart with~$\chi(x_0) \neq 0$. 
The following computation will be carried out in local coordinates. Let~$u_\xi \in \D'(\R)$ be
the family of distributions
\[ u_\xi(g) = \kappa( \chi\, e^{-i \xi x} \otimes g)\:. \]
This family is polynomially bounded in~$\xi$ and, by construction, modulo Schwartz functions in $\xi$ and $\xi_0$, we have
\[ T_t \,u_\xi = e^{i H t} \,u_\xi \:  \mathrm{mod}\, \mathcal{S}_{\xi}(\R^{n-1}). \]
As a consequence,
\[ u_\xi(\eta * g) = \sqrt{2 \pi} \int_0^\infty \hat{\eta}(-\lambda)\: dE_\lambda\, u_\xi(g) \:\mathrm{mod} \,\mathcal{S}_\xi(\R^{n-1}), \]
where~$dE_\lambda$ is the spectral measure of the generator~\eqref{Hdef},
considered as a self-adjoint operator on the Hilbert space~$L^2(\Sigma) \oplus \Omega^1_{(2)}(\Sigma)$.

Choosing a test function~$\eta \in C^\infty_0(\R)$, we have modulo Schwartz functions in $\xi$,
\begin{align*}
\kappa & \Big( \chi(x)\, e^{-i \xi x} \otimes (\eta * \eta)(t) \, e^{-i \xi_0 t} \Big)  \mathrm{mod}\, \mathcal{S}_{\xi,\xi_0}(\R^{n})
= u_\xi \Big( (\eta * \eta) \, e^{-i \xi_0 \cdot} \Big) \\
&= u_\xi \Big( \big(\eta e^{-i \xi_0 \cdot} \big) * \big( \eta e^{-i \xi_0 \cdot} \big) \Big)
= \sqrt{2 \pi} \int_0^\infty \hat{\eta}(-\lambda + \xi_0)\, dE_\lambda\: u_\xi \big(\eta\, e^{-i \xi_0 \cdot}) \big) .
\end{align*}
Taking the Hilbert space norm, one sees that~$\|\eta\, e^{-i \xi_0 \cdot})\|$ is polynomially bounded
in~$(\xi, \xi_0)$, whereas the operator norm of the spectral integral decays rapidly as ~$\xi_0$ goes to~$-\infty$.
We thus obtain rapid decay in~$(\xi, \xi_0)$ in a conic neighborhood of any
direction~$(\tilde{\xi}, \tilde{\xi}_0)$ with~$\tilde{\xi}_0<0$. The result now is implied by the fact that $\kappa$ has its wave-front set contained in the light cone since it solves a homogeneous hyperbolic equation.
\QED
We remark that this result could also be inferred somewhat less directly
from~\cite[Theorem~2.8]{strohmaier+verch}.

\section{Construction of Gupta-Bleuler Representations} \label{sec8}
In this section, we use a  variation of a deformation argument by \cite{fulling+narcowich+wald} to  show that 
Gupta-Bleuler representations and states exist for a large
class of globally hyperbolic space-times. Let~$(M,g)$ be a globally hyperbolic
space-time. According to~\cite{bernal+sanchez}, the manifold admits a
smooth foliation~$(\Sigma_t)_{t \in \R}$ by Cauchy hypersurfaces.
Assume there exists a metric~$g$ on~$\Sigma_0$ such that~$(\Sigma_0, g)$
satisfies condition~(A) in~\eqref{A}.
Then, using the constructions in~\cite{muellero}, there is a globally hyperbolic
space-time~$(\tilde{M}, \tilde{g})$ which is future-isometric to~$(M,g)$
and past isometric to the ultrastatic space-time~$(\R \times \Sigma_0, dt^2 - g)$.
On~$\R \times \Sigma$, we choose~$\kappa$ as in Section~\ref{secultra}.

By propagation of singularities (see~\cite[Theorem~23.2.9]{hormanderIII}) and Lemma~\ref{timeslicelemma}
we have the following.
\begin{Lemma} Suppose that $U$ is an open neighborhood of a Cauchy surface.
Assume that~$\kappa\::\: \Omega^1_0(U) \rightarrow {\mathscr{K}}$ satisfies properties in Section~\ref{sec53} in~$U$, i.e.\
\begin{itemize}
\item[(i)'] $\kappa(\square f) = 0$ for all ~$f \in \Omega^1_0(U)$.
\item[(ii)'] $\la \kappa(f), \kappa(f) \ra \geq 0$ if~$f \in \Omega^1_0(U)$ with~$\delta f=0$.
\item[(iii)'] There is a bilinear form~$G_Z$ on~$\Omega^1_0(U) \times \Omega^1_0(U)$ with smooth integral kernel and the
following properties:
\begin{gather*}
\im \la \kappa(f),\kappa(g) \ra + G_Z(f,g) = G(f,g) \\
\text{The vector space~$Z_U:=\Omega^1_0(U) / \{f \,|\, G_Z(f, \cdot) = 0 \}$
is finite dimensional} \\
G_Z(f, \delta g) = 0 \qquad \text{for all~$f \in \Omega^1_0(U)$ and~$g \in \Omega^2_0(U)$}
\end{gather*}
\item[(iv)'] \hspace*{2.6cm}
$\displaystyle \WF(\kappa) \cap U \subseteq \big\{ (x, \xi) \in \dot{T}^*U \:|\: \xi \in J^+ \big\} \:.$
\item[(v)'] $\mathrm{span}_\mathbb{C} \kappa(\Omega^1_0(U))$ is dense in $\mathscr{K}$.
\item[(vi)'] $\la \kappa(df), \kappa(g) \ra =0$ for all~$f \in C^\infty_0(U)$, $g \in \Omega^1_0(U)$
with~$\delta g=0$ or with $g= dh$ for some $h \in C^\infty_0(M)$.
\end{itemize}
Then there exists a unique extension~$\tilde \kappa\::\: \Omega^1_0(M) \rightarrow {\mathscr{K}}$ that
satisfies (i) everywhere. This unique extension then satisfies (ii),(iii),(iv), (v) and (vi) everywhere.
\end{Lemma} \noindent

One can now construct a Gupta-Bleuler representation on $(\tilde{M}, \tilde{g})$ by constructing the generalized
Fock representation on the ultrastatic space-time~$(\R \times \Sigma_0, dt^2 - g)$. Since there exists a neighborhood $\tilde U_1$
of a Cauchy surface in~$(\tilde{M}, \tilde{g})$ that is isometric to a neighborhood $U_{\mathrm{us}}$
of a Cauchy surface in ~$(\R \times \Sigma_0, dt^2 - g)$ one can use the above Lemma to migrate this Gupta-Bleuler representation
to $(\tilde{M}, \tilde{g})$ by first restricting $\kappa$ to $U_{\mathrm{us}}$ and then extending from $\tilde U$ to~$\tilde M$. 
Similarly, since there exists a neighborhood $U$
of a Cauchy surface in~$(M, g)$ that is isometric to a neighborhood $\tilde U_2$
of a Cauchy surface in~$(\tilde{M}, \tilde{g})$ one constructs a Gupta-Bleuler representation on~$(M, g)$
by restricting $\kappa$ to $\tilde U_2$ and then extending it to $M$.

If a globally hyperbolic space-time has symmetries, one would want to construct Gupta-Bleuler representations
in which these symmetries can be implemented unitarily. Unfortunately, the above method of construction does not respect symmetries. This is of course analogous to the problem of the 
construction of invariant Hadamard states on space-times with non-trivial symmetry group.

\appendix
\section{The Gauge Parameter and Other Commutator Relations} \label{secgaugeparam}
A common procedure in physics is to add a gauge fixing term
to the classical Lagrangian of the electromagnetic field. This leads to the modified
wave equation
\[ \square_\xi A = 0 \qquad \text{where} \qquad
\square_\xi = \delta d + \frac{1}{\xi}\: d \delta \:, \]
which involves the {\em{gauge parameter}} $\xi \in (0, \infty)$.
Choosing~$\xi=1$, the so-called {\em{Feynman gauge}}, gives again the
ordinary wave equation~$\square A=0$. Another common gauge is the
{\em{Landau gauge}} $\xi \searrow 0$ (where the limit is taken after computing expectation values).

Working with a gauge-parameter~$\xi \neq 1$ has the unpleasant consequence that the
modified wave operator~$\Box_\xi$ is not normally hyperbolic. But, in a chosen foliation, 
the modified wave equation can be written as a symmetric hyperbolic system (see for example~\cite{john}),
showing that the Cauchy problem is well-posed, and that the propagation speed is finite.
Moreover, the formulation as a symmetric hyperbolic system depends on the choice of the
foliation.

We now show that working with different gauge parameters gives an equivalent description of the
physical system. To this end, we will construct a bijection of the corresponding field algebras.
In preparation, we relate the solutions of the modified wave equation to the solutions of the ordinary
wave equation: We choose an operator~$R : C^\infty_{\mathrm{sc}}(M) \rightarrow C^\infty_{\mathrm{sc}}(M)$ such that
\[ \Box R f = f \qquad {\text{for all~$f\in C^\infty_{\mathrm{sc}}(M) \cap \ker(\Box_\xi)$}} \:. \]
One method for constructing the operator~$R$ is to solve the Cauchy problem~$\Box \phi = f$
for vanishing initial data on a Cauchy surface.
In Minkowski space, a particular choice of the operator~$R$ is discussed in~\cite{lautrup}
and~\cite[Exercise~7.3]{greiner+reinhardt}. There are of course many other choices.
Suppose that~$\Box_\xi \psi=0$. Then, of course $\Box \delta \psi=0$, and the calculation
\begin{align*}
\Box &\big( \1 + ( \xi^{-1} -1) \, d R \delta \big) \psi
= \big( d \delta + \delta d + (\xi^{-1}-1)\, d \delta d R \delta \big) \psi \\
&= \big( d \delta + \delta d + ( \xi^{-1} -1)\, d \Box R \delta \big) \psi 
= \big( d \delta + \delta d + (\xi^{-1}-1)\, d\delta \big) \psi \\
&= \big( \delta d + \frac{1}{\xi} d \delta \big) \psi =  \Box_\xi \psi = 0
\end{align*}
shows that the following operator maps solutions of the corresponding wave equations into each other,
\[ \I_R := \1 + (\xi^{-1}-1)\, d R \delta \::\: \Omega^1_{\mathrm{sc}}(M) \cap \ker(\Box_\xi) \rightarrow
\Omega^1_{\mathrm{sc}}(M) \cap \ker(\Box) \:. \]
One easily checks that $\I_R$ is invertible with explicit inverse given by
\[\I_R^{-1} = \1 + (1- \xi)\, d R \delta.\]
We thus obtain a one-to-one correspondence
between solutions of the ordinary and modified wave equations.

In order to extend this correspondence to the field operators, we use the following
dual formulation. Assume that~$L$ is a given map~$L : C^\infty_0(M) \rightarrow C^\infty_0(M)$
such that
\[ f - L \Box f \in \im(\Box) \qquad \text{for all~$f \in C^\infty_0(M)$} \:. \]
Again, there are many possibilities to choose~$L$. A particular choice is
\[ L = \eta_- G^0_+ + \eta_+ G^0_- \:, \]
where~$\eta_+$ and~$\eta_-$ are smooth functions with~$\eta_+ + \eta_- = 1$,
having past and future compact support, respectively
(as before,~$G^0_\pm$ denote the causal fundamental solutions for the
scalar wave operator). Then the computation
\[ \big( 1 + (\xi^{-1}-1)\, d L \delta \big) \Box f
= \Box_\xi f + (\xi^{-1} - 1)\, d \big(L \Box -1 \big) \delta f \]
shows that the map
\[ \I_L := 1 + (\xi^{-1} - 1)\, d L \delta \]
has the property
\[ \I_L \Box f \in \im(\Box_\xi) \qquad \text{for all~$f \in \Omega^1_0(M)$} \:. \]
This allows us to identify the field algebras of the modified and the ordinary wave operators via
the relation
\[ \tilde{A}(f) = A(\I_L f) \:. \]
By the above, $\tilde{A}$ satisfies the equation
\[ \Box_\xi \tilde{A} = 0 \]
as an operator-valued distribution. The commutation relations become 
\[ \tilde{A}(f) \tilde{A}(g) - \tilde{A}(f) \tilde{A}(g) = -i \,G(\I_L f , \I_L g) \:. \]

\section{The Construction of States in the BRST Framework} \label{brst}
The purpose of this appendix is to show that a small variation of our construction of Gupta-Bleuler states
can also be used to construct states in the BRST framework. For details of BRST quantization
we would like to refer the reader to~\cite{hollands3}. Here we define only a minimal set of objects
needed for the construction.

As before, let $G^1$ be the difference of advanced and retarded Green's operators on one-forms
and let~$G^0$ be the corresponding operator on functions. We think of $G^1$ and $G^0$ as
bidistributions. Following~\cite[Section 4.2]{hollands3}, one wants to find
bi-distributions $\omega^1 \in (\Omega^1_0(M) \otimes_\pi  \Omega^1_0(M) )'$ and $\omega^0 \in (C^\infty_0(M) \otimes_\pi  C^\infty_0(M) )'$
such that the following list of compatibility conditions is satisfied:
\begin{align*}
\omega^1(\square f,g) =\omega^1( f,\square g) &=0 &&   \forall\: f,g \in \Omega^1_0(M) \\
\omega^0(\square f,g) &=\omega^0( f, \square g) =0 && \forall\: f,g \in C^\infty_0(M) \\
\omega^1(f,g) - \omega^1(g,f) &= -i \;G^1(f,g) &&  \forall\: f,g \in \Omega^1_0(M) \\
\omega^0(f,g) - \omega^0(g,f) &= -i \;G^0(f,g) && \forall\: f,g \in C^\infty_0(M) \\
\omega^0(\delta f, g) &= -\omega^1(f, dg)  && \forall\: f \in \Omega^1_0(M), \:g \in C^\infty_0(M) \\
\omega^1(f,f) &\geq 0 && \forall\: f \in \Omega^1_0(M) \text{ with } \delta f =0 \\
 \omega^0(f,f) &\geq 0 && \forall\: f \in C^\infty_0(M) \:.
\end{align*}
Here we write $\omega(f,g)$ for $\omega(f \otimes g)$, understanding the bidistributions as bilinear forms.
Moreover, one needs to impose the microlocal spectrum condition. 

We will show that one can always construct $\omega^0$ and $\omega^1$ for ultrastatic space-times~$M=\mathbb{R} \times \Sigma$
under the following assumption on $\Sigma$.
\begin{Assumption} \label{assum}
Condition~{\rm{(A)}} in~\eqref{A} holds and $\mathrm{Vol}(\Sigma)=\infty$.
\end{Assumption} \noindent
This assumption implies that the space of square integrable harmonic functions has dimension $0$
but does not rule out the existence of non-trivial square integrable harmonic one-forms. In particular, the assumption is satisfied on compactly supported topological and metric perturbations of $\mathbb{R}^{2n+1}$.
The two point function for one-forms $\omega^1$ will simply be given by
\[ \omega^1(f,g) = \langle \Omega, \hat A(f) \hat A(g) \Omega \rangle, \]
where $\hat A$ is the Gupta-Bleuler representation constructed in Section~\ref{secultra}
and the complex structure is chosen (for the sake of concreteness) as in Corollary \ref{cor53} to be the one associated with the $L^2$-inner product on the 
space of square integrable harmonic forms.

The state $\omega^0$ is obtained if the same construction is carried out on the level of zero forms.
Define the maps $\tau^0$ and $\kappa^0$ by
\begin{align*}
\tau^0 \,:\, \big( \Omega^{0 \perp}_{(2)}(\Sigma)  \big)^2
&\rightarrow {\mathscr{K}}\:, \quad
\begin{pmatrix} \f \\ \dot{\f} \end{pmatrix} \mapsto
\left( \Delta^\frac{1}{4} \f + i \Delta^{-\frac{1}{4}} \dot{\f} \right) \\
\kappa^0 \::\: C^\infty_0(M) &\rightarrow L^2(\Sigma,\C) \:,\quad f \mapsto \tau( \Psi_0^{G^0f})\:,
\end{align*}
where as before we define $\f$ and $\dot \f$ as $\begin{pmatrix} \f \\ \dot{\f} \end{pmatrix} (t)= \Psi^{G f}_t$.
Note that, since square integrable harmonic functions vanish, these do not have to be projected out as it was necessary for forms. In the same way as in Proposition~\ref{prp52}, one shows easily that
the mapping~$\kappa^0$ has the following properties:
\begin{itemize}
\item[(i)] $\kappa^0(\Box f) = 0$ for all~$f \in C^\infty_0(M)$.
\item[(ii)] $\la \kappa^0(f), \kappa^0(f) \ra \geq 0$ for all~$f \in C^\infty_0(M)$.
\item[(iii)] $\im \big( \big\la \kappa^0 (f),\kappa^0 (g) \big\ra \big) = G^0(f,g)$ for all~$f,g \in C^\infty_0(M)$.
\item[(v)] $\mathrm{span}_\mathbb{C} \mathrm{Rg}(\kappa^0)$ is dense in $L^2(\Sigma,\C)$.
\end{itemize}
Moreover, by direct inspection, one can see that
\[ \big( \big\la \kappa^0 (\delta f),\kappa^0 (g) \big\ra \big) = -\big( \big\la \kappa ( f),\kappa (dg) \big\ra \big) \]
for all $f \in \Omega^1_0(M)$ with $\Psi^{G f}_0 \perp \ker \Delta $ and all $g \in C^\infty_0(M)$. 
Indeed,
\begin{align*}
\la \kappa(f), \kappa(dg) \ra &= 
-\big\la \Delta^{\frac{1}{2}} \f_0, \dot \g \big\ra_\Sigma - \big\la \Delta^{-\frac{1}{2}} \dot \f_0, \ddot \g \big\ra_\Sigma
+\big\la \Delta^{\frac{1}{2}}  \f_\Sigma, d_\Sigma \g \big\ra_\Sigma + \big\la \Delta^{-\frac{1}{2}} \dot{\f}_\Sigma,
d_\Sigma \dot{\g} \big\ra_\Sigma\\
&= \big\la \Delta^{\frac{1}{2}} (\dot \f_0 + \delta_\Sigma \f_\Sigma), \g \big \ra_\Sigma + 
\big\la \Delta^{-\frac{1}{2}} (\ddot \f_0 + \delta_\Sigma \dot\f_\Sigma), \dot \g \big \ra_\Sigma=
- \big \la \kappa^0(\delta f), \kappa^0(g) \big \ra \:,
\end{align*}
where we used that $\ddot \f = - \Delta \f$ and $\ddot \g = - \Delta \g$.
This is consistent with the relation~$G(f, dg) = - G^0(\delta f, g)$, which can be verified by a similar computation.
If $\begin{pmatrix} \f \\ \dot{\f} \end{pmatrix} \in \ker \Delta$, then
our assumptions imply that $\f_0 = \dot \f_0 =0$. Since in this case, 
$\f_\Sigma$ and $\dot \f_\Sigma$ are both orthogonal to $d C^\infty_0(\Sigma)$, we infer
that~$\omega^1(f, dg)=0$.
The same argument as in the proof of \ref{prop5.5} shows that $\omega^0$ satisfies the micro local spectrum condition
and we therefore conclude that the two point function $\omega^0$ defined by
\[ \omega^0(f,g):= \big\la \kappa^0 ( f),\kappa^0 (g) \big\ra \]
has all the required properties. By the same argument as in Section \ref{sec8}, it suffices to verify these
properties in a neighborhood of a Cauchy surface. Therefore, the construction extends to all globally
hyperbolic manifolds that admit a Cauchy surface satisfying Assumption~\ref{assum}.

\section{Cohomological Interpretation of the Space~$Z$}\label{cohoint}
We now show that in many situations, the space $Z$ has a cohomological interpretation.
We begin with the case of an ultrastatic manifold~$M = \R \times \Sigma$. 
Then, under the above Assumption~\ref{assum}, the construction in Section~\ref{secultra} can be applied to obtain a generalized Fock representation. In this case, we know that~$H^0_{(2)}=\{ 0 \}$,
and moreover by Corollary~\ref{cor53} the space~$Z$ is naturally isomorphic to~$H^1_{(2)}(\Sigma)
\oplus H^1_{(2)}(\Sigma)$ with standard complex structure. Thus, we obtain the decomposition $Z=Y \oplus \tilde Y$ where
$Y$ is isomorphic to the first direct summand $H^1_{(2)}(\Sigma)$.
Here~$H^p_{(2)}(\Sigma)$ denotes the reduced~$L^2$-cohomology spaces
$\Omega^p_{(2)}(\Sigma) \cap \ker \Delta$. For any~$y \in Y$, there exists a unique
form~$F_y \in \Omega^1(M)$ such that
$$ \la F_y , f \ra = G_Z(\mathfrak{J} y, f ), \quad  \forall\: f \in \Omega^1_0(M) \:. $$
This form is precisely the pull back of the corresponding $L^2$-harmonic form on $\Sigma$ to $\R \times \Sigma$.
Therefore, it not only solves the wave equation but is also closed.
Depending on the interpretation of the reduced $L^2$-cohomology groups on $\Sigma$, this closed form has cohomological significance. For example, if $\Sigma$ is a compactly supported perturbation of $\R^3$, the space $H^1_{(2)}(\Sigma)$ is naturally isomorphic to~$H^1_0(\Sigma,\R)$ so that each $y \not=0$ defines a
non-trivial element in $H^p_0(\Sigma)$.

For general globally hyperbolic $M$ and a given generalized Fock representation, the additional requirement
that the complex structure is chosen such that
$$ G_Z(y, \delta f ) =0 \qquad \forall\: y \in Y, f \in \Omega^2_0(M) $$
seems reasonable and makes sense independent of a choice of foliation by Cauchy surfaces.
For globally hyperbolic manifolds that admit a Cauchy surface satisfying Assumption~\ref{assum},
the deformation argument as in Section \ref{sec8} can be applied to show that such complex structures exist.
Again, since $G_Z$ was assumed to have smooth kernel, for any given $y$ there exists a unique smooth
closed one-form $F_y$ such that
$$  \la F_y , f \ra = G_Z(\mathfrak{J} y , f ) \qquad \forall\: f \in \Omega^1_0(M) \:. $$
Under the above assumptions, this form will be closed and will therefore define a cohomology class.

\Thanks{{{\em{Acknowledgments:}}
We are grateful to Gilles Carron for helpful conversations and to Chris Fewster, Klaus Fredenhagen and
Stefan Hollands for interesting discussions. We would like to thank Claudio Dappiaggi, Andreas Platzer,
Ko Sanders, Alexander Schenkel and Jochen Zahn for valuable comments on the manuscript.
Moreover, we would like to thank the referees for helpful suggestions.
A.S.\ is grateful to the Vielberth foundation, Regensburg, for generous support.}

\providecommand{\bysame}{\leavevmode\hbox to3em{\hrulefill}\thinspace}
\providecommand{\MR}{\relax\ifhmode\unskip\space\fi MR }
\providecommand{\MRhref}[2]{%
  \href{http://www.ams.org/mathscinet-getitem?mr=#1}{#2}
}
\providecommand{\href}[2]{#2}

\end{document}